\documentclass[lettersize,journal]{IEEEtran}
\usepackage[english]{babel}
\usepackage{graphicx}
\usepackage{subfigure}
\usepackage{multicol}
\usepackage{setspace}
\usepackage{amsmath}
\usepackage{amsthm}
\usepackage{epstopdf}
\usepackage{fancyhdr}
\usepackage{indentfirst}
\usepackage{enumerate}
\usepackage{caption}
\usepackage{leftidx}
\usepackage{amssymb}
\usepackage{float}
\usepackage{breqn}
\usepackage{color, xcolor}
\usepackage{bm}
\usepackage{cite}
\usepackage{multirow}
\usepackage{totcount}
\usepackage{algorithm}
\usepackage{algcompatible}
\usepackage{hyperref}
\usepackage{bbm}
\hypersetup{nolinks=true}
\usepackage{chngcntr}
\usepackage{setspace}
\counterwithin*{footnote}{page}

\newtheoremstyle{colon}%
{}
{}
{}
{}
{\itshape}
{:}
{ }
{\thmname{#1}\ \!\thmnumber{\itshape#2}\thmnote{(#3)}}

\newtheoremstyle{colon}%
{}
{}
{\itshape}
{}
{\itshape}
{:}
{ }
{\thmname{#1}\ \!\thmnumber{\itshape#2}\thmnote{(#3)}}

\theoremstyle{colon}
\newtheorem*{Remark*}{Remark}
\newtheorem*{Theorem*}{Theorem}

\newtheorem*{Lemma*}{Lemma}

\newtheorem{Prop}{Proposition}
\numberwithin{Prop}{section}

\numberwithin{Def}{section}

\newtheorem{Rem}{Remark}
\numberwithin{Rem}{section}

\numberwithin{Thm}{section}

\newcommand{\ts}{\textsuperscript}

\begin{document}

\title{Multicast Scheduling over Multiple Channels: A Distribution-Embedding Deep Reinforcement Learning Method}

\author{Ran Li, Chuan Huang, Xiaoqi Qin, Xinyao Nie, and Zhanhong Fu
\thanks{This work was presented in part at the 2022 IEEE/CIC International Conference on Communications in China \cite{iccc}. ({\it Corresponding author: Chuan Huang}.)

R. Li and C. Huang are with the School of Science and Engineering and the Future Network of Intelligence Institute, The Chinese University of Hong Kong, Shenzhen 518172, China, (e-mail: ranli2@link.cuhk.edu.cn; huangchuan@cuhk.edu.cn).

X. Qin is with the State Key Laboratory of Networking and Switching Technology, Beijing University of Posts and Telecommunications, Beijing 100876, China, (e-mail: xiaoqiqin@bupt.edu.cn).

X. Nie and Z. Fu are with SF Technology, Shenzhen 518052, China, (e-mail: sherrynie1@sf-express.com; zhanhongfu@sf-express.com).

}}

\maketitle


\begin{abstract}

Multicasting is an efficient technique for simultaneously transmitting common messages from the base station (BS) to multiple mobile users (MUs). Multicast scheduling over multiple channels, which aims to jointly minimize the energy consumption of the BS and the latency of serving asynchronized requests from the MUs, is formulated as an infinite-horizon Markov decision process (MDP) problem with a large discrete action space, multiple time-varying constraints, and multiple time-invariant constraints. To address these challenges, this paper proposes a novel distribution-embedding multi-agent proximal policy optimization (DE-MAPPO) algorithm, which consists of one modified MAPPO and one distribution-embedding module: The former one handles the large discrete action space and time-varying constraints by modifying the structure of the actor networks and the training kernel of the conventional MAPPO; and the latter one iteratively adjusts the action distribution to satisfy the time-invariant constraints. Moreover, a performance upper bound of the considered MDP is derived by solving a two-step optimization problem. Finally, numerical results demonstrate that our proposed algorithm outperforms the existing ones and achieves comparable performance to the derived benchmark.
\end{abstract}

%
\IEEEpeerreviewmaketitle

\section{Introduction}

The number of mobile users (MUs) and mobile data traffic keep increasing rapidly in recent years. In 2028, the number of MUs is projected to reach 6.16 billion, accounting for approximately 72\% of the world's population \cite{zippia}. Additionally, mobile data traffic in 2028 is projected to reach 329 exabytes per month, which is 3.5-fold of the monthly data traffic in 2022 \cite{ericsson}. However, conventional cellular systems mainly adopt unicast transmissions, where one channel is allocated to one particular MU for data transmissions, and is struggling to cope the aforementioned ever-increasing mobile transmission demand. To address this challenge, it is worth noting that video data traffic, such as short-form videos, advertising videos, and live streaming, is projected to constitute over 70\% of the total mobile data traffic in 2023. Moreover, there is a high degree of content similarity among these video data traffic due to the consistent demand for popular videos from MUs \cite{ericsson}. Leveraging these insights, the adoption of multicast transmission, which allows the base station (BS) to serve multiple MUs with similar data requests over a single channel, has emerged as an effective solution for the future cellular network to accommodate more MUs and handle larger mobile data traffic \cite{david_wc}.

The adoption of multicast technique requires to address two critical challenges. First, multicast transmission might lead to high energy consumption since the transmission power needs to be sufficiently large to serve the MUs with the weakest channel condition, which can be exacerbated when the number of MUs is relatively large. Second, although the requests randomly generated by MUs arrive at the BS asynchronously, they are served by one multicast transmission in a synchronous manner, which may lead to unexpected and unbalanced latency among these requests. Therefore, it is crucial to design scheduling policy over network channels to jointly optimize the average energy consumption and average latency in the multicast networks.

\subsection{Related works}
In recent years, many works have studied the multicast scheduling problem in cellular networks, especially assisted by the advanced techniques such as millimeter wave (mmWave) communication \cite{2020-cap-mmwave}, non-orthogonal multiple access (NOMA) \cite{ee_1}, and intelligent reflecting surface (IRS) \cite{2021-cap-irs1}. One major objective of these works is to maximize the system throughput with limited resources, and various approaches, including deep reinforcement learning (DRL) and conic optimization, have been validated as promising methods to achieve this goal \cite{2020-cap-mmwave, 2021-cap-irs1, 2020-cap-irs, 2020-cap-6g}. Along another research avenue, driven by the ever-increasing attention of green communications \cite{greengood}, the minimization of average energy consumption for multicast transmissions becomes more compelling \cite{ee_1, 2019-ene-mmwave, 2021-ene-mmwave-d2d, ee_4}. To tackle this issue, the authors in \cite{ee_1, 2019-ene-mmwave} designed the optimal precoders at the BS to minimize the average energy consumption for multicast transmissions in the NOMA and mmWave systems, respectively. The authors in \cite{2021-ene-mmwave-d2d} discussed the device-to-device assisted multicast scheduling problem for the mmWave system, and proposed a supervised learning method to minimize the average energy consumption. The authors in \cite{ee_4} studied the joint unicast-multicast scheduling for the NOMA system, and proposed a rate-splitting method to optimize both the spectral efficiency and average energy consumption. Another main challenge in multicast scheduling is the fairness issue, which aims to balance the latencies experienced by different MUs while meeting their various quality of service (QoS) requirements \cite{2021-del-mmwave-ref, 2021-del-mmwave-uav}. However, methods that focus solely on optimizing the QoS metrics tend to prioritize MUs with better channel conditions, resulting in uneven latencies among the MUs with variable channel conditions \cite{ee_1, 2019-ene-mmwave, 2021-ene-mmwave-d2d, ee_4}. To address this issue, the authors in \cite{tradeoff_1} proposed a smart method to optimize the summation of the logarithm QoS metric for all MUs. The logarithm function's ``strictly concave" and ``increasing" properties ensured that MUs with poor channel conditions could be served with reasonable latency. Another approach to address the fairness issue in multicast scheduling was to gather the MUs' requests into request queues \cite{tradeoff_3} and apply Lyapunov optimization to ensure the rate stability for these queues, which guaranteed the request latencies to be finite. However, the proposed algorithms in \cite{tradeoff_1, tradeoff_3} only provided a rough control on the latency, and achieving the fine-grained optimization, such as minimizing the latency and simultaneously satisfying the QoS requirements, was proved to be out the scope of these algorithms.

To jointly optimize the average energy consumption and the average latency, one widely adopted approach is to minimize the weighted summation of them \cite{huang_stop, cui_ite, 2020-joint, mdp_2}, which is typically formulated as an infinite-horizon Markov decision process (MDP) and is difficult to be solved due to ``the curse of dimensionality" \cite{bertsekas_dp}. The authors in \cite{huang_stop, cui_ite, 2020-joint, mdp_2} proposed some efficient while suboptimal methods to deal with this challenge under various special scenarios. Specifically, the authors in \cite{huang_stop} studied the multicast scheduling problem where multiple MUs request one common message from the BS and the MUs have mixed latency penalties, and a suboptimal scheduling policy was developed based on the optimal stopping theorem. The authors in \cite{cui_ite} discussed the multicast scheduling problem with multiple requested messages and single available channel, and the optimal scheduling policy was constructed based on the classical relative value iteration (RVI) algorithm. The authors in \cite{2020-joint, mdp_2} deployed DRL algorithms to study the joint multicast scheduling and message caching problems in the ultra dense networks and the heterogeneous networks, respectively, and again focused on the scenarios with multiple requested messages and single available channel. Notably, the authors in \cite{huang_stop, cui_ite, 2020-joint, mdp_2} assumed that the multicasting of each message on each channel starts and ends synchronously, i.e., all channels are always released from multicastings simultaneously and available at every scheduling moment.

To the best of our knowledge, these is no existing work discussed the scheduling problem for asynchronous multicasting, where the multicast transmissions of different messages over different channels consume different time durations, resulting in asynchronous starting and ending times for different multicast transmissions.

\subsection{Main contributions}

This paper studies a general time-slotted multicast scheduling problem in cellular networks with one BS and multiple MUs. Specifically, the BS employs caching techniques to store multiple messages of various sizes \cite{2020-joint, mdp_2}, and the MUs randomly send request for downloading these messages from the BS. To serve these MUs, the BS utilizes multiple channels to multicast the messages, and adopts a multicast scheduling policy to determine which message is multicasted over each channel. Furthermore, we take into account the asynchronous nature of multicast transmission, where the multicasting of different messages over different channels incurs various transmission durations. The objective of this paper is to develop a multicast scheduling policy at the BS that jointly optimizes the average energy consumption and average latency penalty in the cellular network. We summarize our contributions as follows:
\begin{itemize}
    \item We analyze the dynamics of the multicast scheduling problem and derive the formulations for both the energy consumption and latency penalty. We also formulate two series of constraints inherent to the problem: the first series enforces the occupation rule for channel allocation, and the second series regulates the multicasting rule for message transmission. Then, we formulate the multicast scheduling problem as an infinite-horizon MDP, which is challenging due to the large discrete action space, multiple time-varying constraints, and multiple time-invariant constraints. 
    \item To address the above challenges, we convert the MDP problem into an equivalent Markov game and propose the distribution-embedding multi-agent proximal policy optimization (DE-MAPPO) to efficiently solve it, which consists of a modified MAPPO and a distribution-embedding module. The modified MAPPO utilized adapted actor networks and an enhanced method for calculating probability ratio during offline training to tackle the challenges of large discrete action spaces and time-varying constraints. The proposed distribution-embedding (DE) module modifies the output distribution of MAPPO iteratively to address the time-invariant issue and achieves better convergence performance than the conventional action-embedding approach.
	\item We derive an upper bound for the performance of the considered MDP, serving as a benchmark to evaluate the proposed DE-MAPPO. First, we  simplify the MDP by relaxing its time-varying and time-invariant constraints and recasting it as a two-step optimization problem. Then, we utilize either vanilla deep Q network (DQN) or optimal stopping technique to solve the first-step optimization, and employ either integer programming or the Lagrange multiplier technique to solve the second-step one.
\end{itemize}

\begin{figure*}
\centering
\includegraphics[width=6.8in]{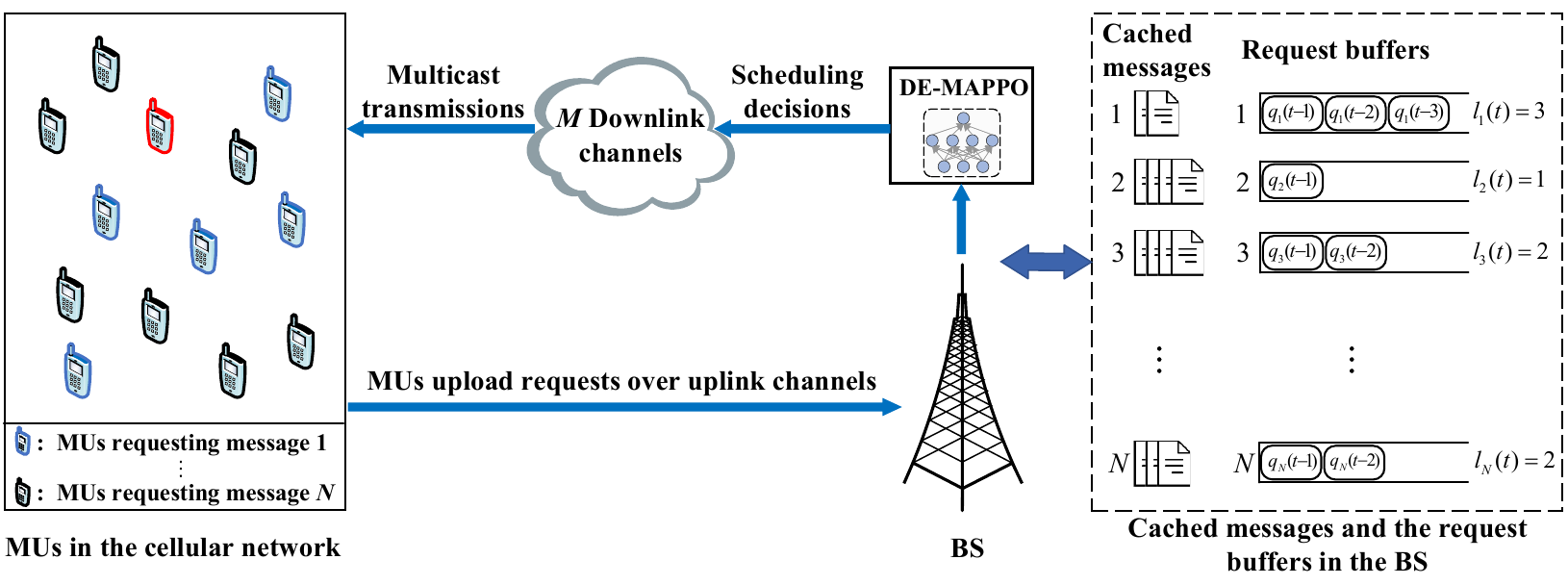}
\caption{Multicast scheduling over multiple channels.}\label{fig:system}
\end{figure*}

The remainder of this paper is organized as follows. Section \ref{SectionII} introduces the system model and formulates the multicast scheduling problem. Section \ref{SectionIII} presents the proposed algorithm. Section \ref{SectionIV} establishes an upper bound for the multicast scheduling problem as benchmark. Section \ref{SectionV} presents the simulation results of the proposed algorithm. Finally, Section \ref{SectionVI} concludes this paper.

\section{System Model and Problem Formulation}\label{SectionII}
Consider a time-slotted cellular network as depicted in Fig.~\ref{fig:system}, which consists of a large number of MUs, one BS caching with $N$ messages, and $M$ available channels for multicast transmissions. During each time slot, the MUs send requests for downloading one or multiple cached messages from the BS. Then, the BS catalogs these requests based on their requested messages and stores them in $N$ request buffers. At the end of each time slot, the BS collects the statuses of all $M$ channels and $N$ request buffers, and uses this information to determine the selection of messages and the allocation of channels for multicast transmissions. 

To provide a clear description of the multicast scheduling for the considered cellular network, we adopt a MDP framework, which comprises four fundamental components: state, action, transitions, and reward.

\subsubsection{State}
The state of the considered system contains the request matrix, channel availability vector, and channel status matrix.

{\textbf{Request matrix:}} Denote the number of the requests for the $n$\ts{th} message arriving within the $t$\ts{th} time slot as $q_n(t)$ and the number of time slots after the previous multicasting of the $n$\ts{th} message as $l_n(t)$. Here, $q_n(t)$ is modeled as an independent and identically distributed (i.i.d.) random variable across $t$, with its mean being $\lambda_n$ and its probability mass function (pmf) being $f_n$. Then, within the past $l_n(t)$ time slots, all the arrived requests for the $n$\ts{th} message are stored at the $n$\ts{th} request buffer and denoted as vector $\bm{q}_n(t)$, i.e.,
\begin{align}\label{def:qn}
	\bm{q}_n(t)\!\triangleq\!\!\left\{\!\!\begin{array}{ll}
		\!\!\Big[q_n(t-1),\cdots,q_n\left(t-l_n(t)\right),&\\
		\underbrace{0,\cdots,0}_{M^*-l_n(t)\ \text{zeros}}\Big]^T&l_n(t)\!\leq\!M^*\\
		\!\!\Big[q_n(t-1),\cdots,q_n(t\!-\!M^*\!+\!1),&\\
		\sum_{\tau=M^*}^{l_n(t)}q_n(t-\tau)\Big]^T&l_n(t)\!>\!M^*,
	\end{array}\right.
\end{align}
where $M^*$ is the size of the request buffer. If $l_n(t)>M^*$, the number of all the requests arrived before the $(t-M^*)$\ts{th} time slot is added into the last entry of $\bm{q}_n(t)$. Finally, the requests for all messages are denoted as request matrix $\bm{Q}(t)\!\!\triangleq\!\!\left[\bm{q}_1(t),\bm{q}_2(t),\!\cdots,\!\bm{q}_N(t)\right]^T$\!\!, which has the dimensionality of $N$-by-$M^*$.

{\textbf{Channel availability vector:}} We consider the scenario that each multicast transmission of the $n$\ts{th} message over the $m$\ts{th} channel consumes $T_{n,m}\in\mathbb{Z}^+$ consecutive time slots. Then, during these $T_{n,m}$ slots, the $m$\ts{th} channel is not available for new multicasting. To track the availability of the $m$\ts{th} channel, we define $c_m(t)$ as a channel indicator: If the $m$\ts{th} channel is neither scheduled nor reserved for a previous multicast transmission, $c_m(t)$ is set to $0$; otherwise, $c_m(t)$ is set as the number of remaining time slots before the $m$\ts{th} channel is released. Finally, we obtain the channel availability vector at the $t$\ts{th} time slot as $\bm{c}(t)=[c_1(t),c_2(t),\cdots,c_M(t)]^T$.

{\textbf{Channel status matrix:}} The downlink channel power gain from the BS to the $q$\ts{th} MU that has requested the $n$\ts{th} message within the $t$\ts{th} time slot over the $m$\ts{th} channel is denoted as $g_{n,m,q}(t)$, which is considered to be i.i.d. across different $q$ and $t$. Note that to successfully multicast the $n$\ts{th} message over the $m$\ts{th} channel, the energy consumption at the BS is determined by the worst channel power gain among the MUs whose requests are stored in $\bm{q}_n(t)$, which is given as
\begin{align}\label{def:gnm}
\begin{split}
g_{n,m}(t)\!\triangleq\!\min_{\tau\in\{1,\cdots,M^*\}}\min_{q\in\{1,\cdots,q_n(t-\tau)\}}g_{n,m,q}(t\!-\!\tau).
\end{split}
\end{align} 
Finally, we define the channel status matrix at the $t$\ts{th} time slot as $\bm{G}(t)\in\mathbb{R}_{>0}^{N\times M}$, with the $(n,m)$\ts{th} entry of $\bm{G}(t)$ being $g_{n,m}(t)$, i.e., $[\bm{G}(t)]_{(n,m)}\triangleq g_{n,m}(t)$. Here, $\mathbb{R}_{>0}$ is the set of all positive real numbers.

Therefore, the state of the considered system at the $t$\ts{th} time slot can be represented by the triple $\bm{s}(t)\triangleq(\bm{Q}(t),\bm{c}(t),\bm{G}(t))$ with dimensionality of $NM^*+M+NM$.

\subsubsection{Action}
Denote the scheduling decision for the $m$\ts{th} channel at the $t$\ts{th} time slot as $a_m(t)\in\{0,1,\cdots,N\}$: $a_m(t)=0$ indicates that the $m$\ts{th} channel is not scheduled to multicast any message; otherwise, it is scheduled to multicast the $a_m(t)$\ts{th} message. Then, define the action of the considered system as $\bm{a}(t)\triangleq[a_1(t),a_2(t),\cdots,a_M(t)]^T$. Note that if the $m$\ts{th} channel has been reserved for multicasting in previous slots, i.e., $c_{m}(t)>0$, it cannot be scheduled to multicast any new message, which implies
\begin{align}\label{con:cmc1}
    c_{m}(t)a_m(t)=0,\ m\in\mathcal{M}.
\end{align}
Here, $\mathcal{M}\triangleq\left\{1,2,\cdots,M\right\}$. Furthermore, we consider the scenario that each message cannot be simultaneously multicasted over multiple channels, which is a common restriction in existing multicast protocols \cite{new_cite1,new_cite2,new_cite3} to prevent wastage of channel resources and transmission energy. Then, this restriction is formulated as
\begin{align}\label{con:cmc2}
a_m(t)\neq a_{m'}(t),\ \text{if}\ a_m(t)>0\ \text{and}\ m\neq m'.
\end{align}

\subsubsection{Transitions}\label{section_trans}
The request matrix, channel availability vector, and channel status matrix are updated over time according to the scheduling decisions for $M$ channels. We denote $b_n(t)$ as an indicator on whether the $n$\ts{th} message is multicasted at the $t$\ts{th} time slot, i.e.,
\begin{align}\label{def:bn}
	b_n(t)\triangleq\mathcal{I}\left(\sum_{m=1}^M\mathcal{I}_n\left(a_m(t)\right)\right), n\in\mathcal{N},
\end{align}
where $\mathcal{N}$ is defined as $\mathcal{N}\triangleq\{1,2,\cdots,N\}$; $\mathcal{I}_n(x)$ equals 1 if $x$ is $n$, and equals 0 otherwise; and $\mathcal{I}(x)$ equals 1 if $x$ is positive, and equals 0 otherwise. Apparently, if the $n$\ts{th} message is multicasted over some channels at the $t$\ts{th} time slot, $b_n(t)=1$; otherwise, $b_n(t)=0$. Then, the transitions of the three aforementioned entities are derived as follows.

{\textbf{Request matrix:}} We update the request matrix in the following two cases:
\begin{itemize}
	\item If the $n$\ts{th} message is selected for multicasting at the $t$\ts{th} time slot, i.e., $b_n(t)=1$, all the previously stored requests in the $n$\ts{th} request buffer will be served and only $q_n(t)$ new requests that arrive within the $t$\ts{th} time slot will be stored in the first entry of the $n$\ts{th} request buffer;
	\item If the $n$\ts{th} message is not selected for multicasting, the previously stored requests in the $n$\ts{th} request buffer will be delayed for one more time slot, and thus all the entries of $\bm{q}_n(t)$ will move one-entry rightward. Meanwhile, $q_n(t)$ new requests will be stored in the first entry of the $n$\ts{th} request buffer.
\end{itemize}
To summary, we have
\begin{align}\label{eq:qn}
	\bm{q}_n(t+1)\!=\!\!\left\{\!\!\begin{array}{ll}
	    \!\Big[q_n(t),\underbrace{0,\cdots,0}_{M^*-1\ \text{zeros}}\Big]^T&b_n(t)\!=\!1\\
        \!\Big[q_n(t),[\bm{q}_n(t)^T]_{(1:M^*-2)},&\\
        \!\big[\bm{q}_n(t)\big]_{M^*-1}+[\bm{q}_n(t)]_{M^*}\Big]^T&b_n(t)\!=\!0,
	\end{array}\right.
\end{align}
for all $n\in\mathcal{N}$, where $[\bm{q}_n(t)^T]_{1:M^*-2}$ is a $1$-by-$(M^*-2)$-dimension vector containing the first to the $(M^*-2)$\ts{th} entries of $\bm{q}_n(t)^T$.

{\textbf{Channel availability vector:}} We update the channel availability vector by the following three cases:
\begin{itemize}
	\item If the $m$\ts{th} channel has been reserved for multicasting the $n$\ts{th} message in previous slots, i.e., $c_{m}(t)>0$. The remaining time for the release of the $m$\ts{th} channel decreases by one, i.e., $c_{m}(t+1)=c_{m}(t)-1$;
	\item If the $m$\ts{th} channel is not reserved but selected for multicasting the $n$\ts{th} message at the $t$\ts{th} time slot, i.e., $c_{m}(t)=0$ and $a_m(t)=n$, it will be released after $T_{n,m}-1$ time slots, i.e., $c_{m}(t+1)=T_{n,m}-1$;
	\item If the $m$\ts{th} channel is neither reserved nor selected for new multicasting at the $t$\ts{th} time slot, i.e., $c_{m}(t)=0$ and $a_m(t)=0$, then it follows $c_{m}(t+1)=0$.
\end{itemize}
To summary, we have
\begin{align}\label{eq:cnm}
    c_{m}(t+1)\!=\!\left\{\!\!\begin{array}{ll}
        \!c_{m}(t)\!-\!1 &c_{m}(t)>0\\
        \!T_{a_m(t),m}-1 &c_{m}(t)=0, a_m(t)>0\\
        \!0 &c_{m}(t)=0, a_m(t)=0.
    \end{array}\right.
\end{align}

{\textbf{Channel status matrix:}} We update the channel status matrix by the following two cases:
\begin{itemize}
	\item If the $n$\ts{th} message is selected for multicasting at the $t$\ts{th} time slot, i.e., $b_n(t)=1$, then based on \eqref{def:qn}, we have $\bm{q}_n(t+1)=[q_n(t),0,\cdots,0]^T$. By taking $\bm{q}_n(t+1)$ into the definition in \eqref{def:gnm}, we have $g_{n,m}(t+1)=\min_{q\in\{1,\cdots,q_n(t)\}}g_{n,m,q}(t)$;
	\item If the $n$\ts{th} message is not selected for multicasting, i.e., $b_n(t)=0$, then similarly, we have $\bm{q}_n(t+1)$ $=\!\Big[q_n(t),[\bm{q}_n(t)^T]_{(1:M^*-2)},\big[\bm{q}_n(t)\big]_{M^*-1}\!+\![\bm{q}_n(t)]_{M^*}\Big]^T$ and $g_{n,m}(t+1)=\min\{g_{n,m}(t),\min_{q\in\{1,\cdots,q_n(t)\}}g_{n,m,q}(t)\}$.
\end{itemize}
To summary, we have
\begin{align}\label{eq:gnm}
\begin{split}
g_{n,m}(t\!+\!1)\!=\!\left\{\begin{array}{ll}
\!\min_{q\in\{1,\cdots,q_n(t)\}}g_{n,m,q}(t)&\!b_n(t)\!\!=\!\!1\\
\!\min\{g_{n,m}(t),&\\
\!\min_{q\in\{1,\cdots,q_n(t)\}}g_{n,m,q}(t)\}&\!b_n(t)\!\!=\!\!0.
\end{array}\right.
\end{split}
\end{align}

\subsubsection{Reward}\label{sectionreward}
This paper aims to optimize both the average energy consumption and average latency penalty for the multicast scheduling system. Therefore, we use the energy consumption per time slot and latency penalty per time slot to formulate the system reward at each time slot.

{\textbf{Energy consumption:}} The energy consumption of multicasting the $n$\ts{th} message over the $m$\ts{th} channel within the $t$\ts{th} time slot is $\frac{Z_{n,m}}{g_{n,m}(t)}$ \cite{david_wc}, where $Z_{n,m}=T_0\left(2^{\frac{R_{n}}{B_mT_{n,m}T_0}}-1\right)$ is a constant, with $T_0$, $R_{n}$, and $B_m$ being the duration of one time slot, the number of information bits of the $n$\ts{th} message, and the bandwidth of the $m$\ts{th} channel, respectively. Then, the total energy consumption of the multicast transmissions taking place over $M$ channels and starting from the $t$\ts{th} time slot is computed as 
\begin{align}\label{def:obj1}
\sum_{n=1}^N\sum_{m=1}^{M}\frac{T_{n,m}Z_{n,m}}{g_{n,m}(t)}\mathcal{I}_n\left(a_m(t)\right).
\end{align}

{\textbf{Latency penalty:}} At the $t$\ts{th} time slot, the $n$\ts{th} request buffer stores $[\bm{q}_n(t)]_1$, $[\bm{q}_n(t)]_2$, $\cdots$, and $[\bm{q}_n(t)]_{M^*}$ requests that have been delayed for $1$, $2$, $\cdots$, and $M^*$ time slots, respectively. Denote the penalty for delaying one request for the $n$\ts{th} message by $\tau$ time slots as $p_n(\tau)$. The requests stored in the $n$\ts{th} request buffer contribute a latency penalty of $[\bm{q}_n(t)]_1p_n(1)+[\bm{q}_n(t)]_2p_n(2)+\cdots+[\bm{q}_n(t)]_{M^*}p_n(M^*)=\sum_{\tau=1}^{M^*}[\bm{q}_n(t)]_{\tau}p_n(\tau)$ at the $t$\ts{th} time slot. Then, the total latency penalty incurred by the requests stored in $N$ request buffers at the $t$\ts{th} time slot is calculated as
\begin{align}\label{def:obj2}
\sum_{n=1}^{N}\sum_{\tau=1}^{M^*}[\bm{q}_n(t)]_{\tau}p_n(\tau).	
\end{align}

The reward $r(t)$ of the considered system is defined as the weighted sum of the energy consumption in \eqref{def:obj1} and the latency penalty in \eqref{def:obj2}, i.e.,
\begin{align}\label{def:reward}
\begin{split}
	r(t)\!\triangleq\!&-\!\!\Big(\!V\sum_{n=1}^N\sum_{m=1}^{M}\frac{T_{n,m}Z_{n,m}}{g_{n,m}(t)}\mathcal{I}_n\left(a_m(t)\right)\\
	&+\sum_{n=1}^{N}\sum_{\tau=1}^{M^*}[\bm{q}_n(t)]_{\tau}p_n(\tau)\Big),
\end{split}
\end{align}
where $V>0$ is the tradeoff parameter.

\subsubsection{Problem formulation}
We aim to maximize the long-term average reward under various constraints and thus formulate the multicast scheduling problem as
\begin{align}
\textbf{(P1)}\ &\max_{\left\{\bm{a}(t)\right\}}\ \lim_{T\to\infty}\mathbb{E}_{\{q_n(t)\},\{g_{n,m,q}(t)\}}\left[\frac{1}{T}\sum_{t=1}^{T}r(t)\right] \label{obj:p1}\\
&\ \ \text{s.t.}\ \ \ \eqref{con:cmc1}, \eqref{con:cmc2}, \eqref{eq:qn},\eqref{eq:cnm},\eqref{eq:gnm}.\nonumber
\end{align}

\begin{Rem}\label{rem_21}
From \eqref{obj:p1}, it is observed that there are three challenges in Problem {\normalfont\bf{(P1)}}: a large state space when $N$, $M$, or $M^*$ is large, a large discrete action space when $N$ or $M$ is large, and the coexistence of time-varying and time-invariant constraints in \eqref{con:cmc1} and \eqref{con:cmc2}.
\begin{itemize}
\item The first challenge cannot be addressed by conventional dynamic programming, since solving the Bellman optimality equation requires significant computational resources when the state space is large {\normalfont\cite{bertsekas_dp}}.
\item Modern deep reinforcement learning (DRL) algorithms cannot efficiently solve the MDPs with large discrete action space since they require a policy network of prohibitively large scale to handle such action space, resulting in difficulties in achieving convergence. To alleviate this challenge, existing works proposed three techniques: (1) The first one is to modify the MDP by converting its original action to a one-hot action {\normalfont \cite{dqn,ddpg}}. However, as the dimension or domain of the original action space increases, this method leads to an explosion of the dimensionality of the one-hot vector. (2) The second one, known as action-embedding {\normalfont \cite{wolper,yingjun}}, uses deep deterministic policy gradient (DDPG) to output a continuous-valued action with the same dimensionality as the original action space. Then, this approach selects several discrete-valued actions near the obtained continuous-valued one as candidate actions and picks the best-evaluated candidate action as the final action. However, this approach may suffer from unstable convergence and poor convergence performance when the action space is large {\normalfont\cite{wolper}}. (3) The last approach is multi-agent DRL (MADRL) {\normalfont\cite{sutton_rl}}, which uses individual policy networks to determine the action value on each dimension. This approach has promising convergence performance and does not suffer from the large discrete action space issue. However, it cannot deal with the MDPs with time-varying or time-invariant constraints.
\item Both the time-varying and time-invariant constraints are difficult to handle for DRL or MADRL algorithms {\normalfont\cite{sutton_rl}}, as their trial-and-error mechanism makes it almost unavoidable to visit infeasible actions. In {\normalfont\cite{liran1}}, the authors proposed a novel MADRL-based algorithm, which modifies the policy network structure and can efficiently solve MDPs with time-varying constraints. However, this algorithm is still not applicable for solving MDPs with a combination of time-varying and time-invariant constraints.
\end{itemize}
\end{Rem}

To address the aforementioned challenges, we transform problem {\bf{(P1)}} into an equivalent Markov game {\bf{(P2)}} by treating each channel as an individual agent. Problem {\bf{(P2)}} is specified as follows:
\begin{itemize}
\item {\bf Agent observation:} The $m$\ts{th} agent's observation is denoted as $\bm{s}_m(t)\triangleq(\bm{Q}(t),c_m(t),\bm{g}_m(t))$, where $\bm{g}_m(t)$ represents the channel power gains for multicasting $M$ messages over the $m$\ts{th} channel and is defined as $\bm{g}_m(t)\triangleq[g_{1,m}(t),\cdots,g_{N,m}(t)]^T$;
\item {\bf Agent action:} The action of the $m$\ts{th} agent is $a_m(t)$, which is also referred to as the $m$\ts{th} agent action;
\item {\bf Agent reward:} The reward for each agent is $r(t)$.
\end{itemize}
The equivalence between {\bf{(P1)}} and {\bf{(P2)}} can be proven using similar method as that in \cite{liran1} and thus is omitted here for simplicity. To handle the coexistence of time-varying and time-invariant constraints, we propose a modified MADRL algorithm, which will be discussed in the next section.

\begin{figure*}[!htb]
\centering
\includegraphics[width=5.6in]{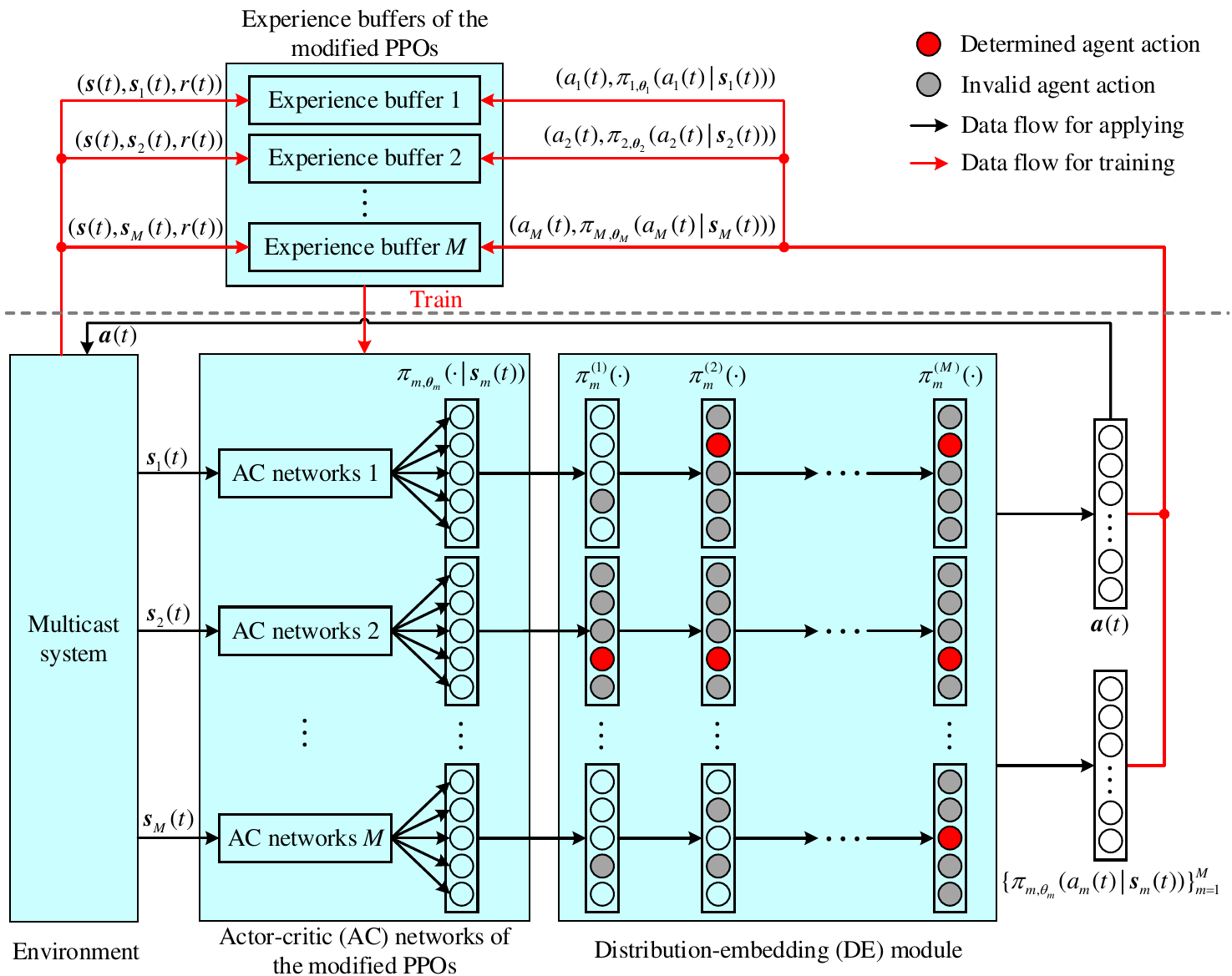}
\caption{Structure of the proposed DE-MAPPO.}\label{fig:alg}
\end{figure*}

\section{Distribution-Embedding Multi-Agent Proximal Policy Optimization}\label{SectionIII}
In this section, we propose a DE-MAPPO algorithm to solve problem {\bf{(P2)}}, which enables efficient multicast scheduling while preserving the satisfaction of both the time-varying and time-invariant constraints. Specifically, we first describe the structure of the proposed algorithm. Then, we present the offline training algorithm and the online applying algorithm to solve problem {\bf (P2)}.

\subsection{Structure of proposed algorithm}\label{algorithm}
The proposed algorithm consists of $M$ modified PPO modules and a DE module.
\subsubsection{Modified PPOs}
The $m$\ts{th} modified PPO is responsible for determining the $m$\ts{th} agent action $a_m(t)$ and is denoted as PPO$_m$. As illustrated in Fig.~\ref{fig:alg}, PPO$_m$ contains an actor-critic (AC) network and an experience buffer, and the AC network consists of an actor network and a critic network.
\begin{itemize}
    \item {\bf Actor network:} The actor network of PPO$_m$ is a fully connected neural network (NN) denoted as $\pi_{m,\bm{\theta}_m}$, with $\bm{\theta}_m$ representing the embedded parameters. This actor network takes $\bm{s}_m(t)$ as input and outputs the probabilities of assigning $a_m(t)$ with each possible value in $\{0,1,\cdots,N\}$. These probabilities and their corresponding distribution are denoted as $\{\pi_{m,\bm{\theta}_m}(a|\bm{s}_m(t))\}_{a=0}^{N}$, and $\pi_{m,\bm{\theta}_m}(\cdot|\bm{s}_m(t))$, respectively. However, when the $m$\ts{th} channel is currently unavailable, i.e., $c_m(t)>0$, the above distribution $\pi_{m,\bm{\theta}_m}(\cdot|\bm{s}_m(t))$ cannot guarantee $a_m(t)$ to be assigned with $0$, and thus violates the time-varying constraints in \eqref{con:cmc1}. To address this challenge, we incorporate the methodology proposed in \cite{liran1} and adjust the distribution $\pi_{m,\bm{\theta}_m}(\cdot|\bm{s}_m(t))$ in the case with $c_m(t)>0$ as
    \begin{align}\label{eq:ppo_mod}
    \pi_{m,\bm{\theta}_m}(a|\bm{s}_m(t))=\left\{\begin{array}{ll}
	    1&a=0\ \text{and}\ c_m(t)>0\\
	    0&a>0\ \text{and}\ c_m(t)>0.
        \end{array}\right.
    \end{align}
    With this modification,  $a_m(t)$ will unequivocally adhere to the time-varying constraints highlighted in \eqref{con:cmc1}.
    \item {\bf Critic network:} The critic network of PPO$_m$ is also a fully connected NN denoted as $V_{m,\bm{\phi}_m}$, with $\bm{\phi}_m$ representing the embedded parameters. This critic network takes $\bm{s}(t)$ as input and outputs the value function (as defined in \cite{sutton_rl}) of the state $\bm{s}(t)$, which is denoted as $V_{m,\bm{\phi}_m}(\bm{s}(t))$;
    \item {\bf Experience buffer:} The experience buffer of PPO$_m$ stores the generated experiences during the offline training phase, which are the five-component tuples $(\bm{s}(t),\bm{s}_m(t),r(t),a_m(t),$ $\pi_{m,\bm{\theta}_m}(a_m(t)|\bm{s}_m(t)))$.
\end{itemize}

\subsubsection{DE module}
The DE module employs the distributions generated by $M$ modified PPOs, i.e., $\{\pi_{m,\bm{\theta}_m}(\cdot|\bm{s}_m(t))\}_{m=1}^M$, to determine the action $\bm{a}(t)$ through the following steps: We first carry out the stochastic agent ordering step; then, we perform agent action selection and distribution modification iteratively for $M$ rounds; finally, we proceed with the agent action combination step.
\begin{itemize}
\item {\bf Stochastic agent ordering:} In this step, we determine the sequence in which each agent selects its action. First, we randomly permute vector $[1,2,\cdots,M]^T$ to generate a random permutation $\bm{u}(t)\in\{1,2,\cdots,M\}^{M\times 1}$, i.e.,
\begin{align}\label{eq:de1}
\bm{u}(t)=\text{Perm}([1,2,\cdots,M]^T),
\end{align}
where $\text{Perm}(\bm{x})$ is a function to generate a random permutation of vector $\bm{x}$. Then, the elements of $\bm{u}(t)$ determine the order in which each agent takes action. Specifically, we denote the $m$\ts{th} element of $\bm{u}(t)$ as $u_m(t)$. Then, the $u_m(t)$\ts{th} agent is the $m$\ts{th} one to determine its action.

\item {\bf $m$\ts{th} agent action selection:} In this step, we determine $a_{u_m(t)}(t)$ by sampling a random value from distribution $\pi^{(m-1)}_{u_m(t)}(\cdot)$, i.e.,
\begin{align}\label{eq:de2}
\text{Pr}\{a_{u_m(t)}(t)\!\!=\!\!a\}\!=\!\pi^{(m-1)}_{u_m(t)}(a),\forall a\!\in\!\{0,1,\cdots,N\},
\end{align}
where the distributions $\{\pi^{(m-1)}_{m'}(\cdot)\}_{m'=1}^M$ are generated from the previous iteration of the distribution modification, as introduced in the next step. Moreover, the initial distributions are defined as  $\pi^{(0)}_{m'}(\cdot)\triangleq\pi_{m',\bm{\theta}_{m'}}(\cdot|\bm{s}_{m'}(t))$ for all $m'\in\mathcal{M}$.

\item {\bf $m$\ts{th} distribution modification:} In this step, since the $u_1(t)$\ts{th} to the $u_m(t)$\ts{th} agents have already determined their actions, we focus on generating new distributions $\pi^{(m)}_{m'}(\cdot)$ for the remaining agents with indices in $\{u_{\hat{m}}(t)|\hat{m}\in\mathcal{M};\hat{m}>m\}$. Specifically, we aim to ensure that the new distribution $\pi^{(m)}_{m'}(\cdot)$ maintains similarity with the one generated in the previous distribution modification, i.e., $\pi^{(m-1)}_{m'}(\cdot)$. To achieve this goal, we introduce a temporary distribution $\hat{\pi}_{m'}(\cdot)$ to represent the new distribution and initialize it as
\begin{align}\label{eq:de3}
	\hat{\pi}_{m'}(\cdot)=\pi^{(m-1)}_{m'}(\cdot).
\end{align}
Then, to avoid violating the time-invariant constraints in \eqref{con:cmc2}, we must ensure that the $(m')$\ts{th} agent does not assign the action $a_{u_m(t)}(t)$ of the $u_m(t)$\ts{th} agent to its own action. To achieve this goal, we modify the temporary distribution as
\begin{align}\label{anti:con}
	\hat{\pi}_{m'}(a_{u_m(t)}(t))=0,\ \text{if}\ a_{u_m(t)}(t)\neq 0.
\end{align}
Finally, we normalize the temporary distribution $\hat{\pi}_{m'}(\cdot)$ and assign the result to $\pi^{(m)}_{m'}(\cdot)$, i.e.,
\begin{align}\label{eq:de4}
	\pi^{(m)}_{m'}(a)=\frac{\hat{\pi}_{m'}(a)}{\sum_{a=0}^N\hat{\pi}_{m'}(a)},\forall a\in\{0,1,\cdots,N\}.
\end{align}

\item {\bf Agent actions combination:} In this step, we combine the agent actions selected in $M$ iterations of the agent action selections to generate action $\bm{a}(t)$, where $a_{u_m(t)}(t)$ with $u_m(t)\in\mathcal{M}$ is obtained in the $m$\ts{th} iteration of agent action selections according to \eqref{eq:de2}. Additionally, we collect the values of $\{\pi_{m,\bm{\theta}_m}(a_m(t)|\bm{s}_m(t))\}_{m=1}^M$ for offline training.
\end{itemize}

\begin{Rem}
Based on \eqref{eq:de3}, \eqref{anti:con}, and \eqref{eq:de4}, the distributions generated during the $m$\ts{th} distribution modification avoid assigning the value $a_{u_m(t)}(t)$ to the actions of the remaining agents with indices in $\{u_{\hat{m}}(t)|\hat{m}\in\mathcal{M};\hat{m}>m\}$. By implementing this step for $M$ iterations, all the values assigned to the $M$ agents are distinct from each other, ensuring that the resulting action $\bm{a}(t)$ satisfies the time-invariant constraint in \eqref{con:cmc2}. Additionally, based on \eqref{eq:de2}, \eqref{eq:de3}, \eqref{anti:con}, and \eqref{eq:de4}, we observe that $a_{u_m(t)}(t)$, $m\in\mathcal{M}$, is sampled from the distribution $\pi^{(m-1)}_{u_m(t)}(\cdot)$, and this distribution is derived through $m-1$ iterations of modifications to $\pi_{u_m(t),\bm{\theta}_{u_m(t)}}(\cdot|\bm{s}_{u_m(t)}(t))$. Consequently, the distribution of $a_m(t)$ closely resembles $\pi_{m,\bm{\theta}_m}(\cdot|\bm{s}_m(t))$, and this resemblance diminishes as indicated by the order of indices in vector $\bm{u}(t)$. 

In conclusion, the DE module ensures that not only the generated action $\bm{a}(t)$ adheres to the time-invariant constraints in \eqref{con:cmc2}, but each element of $\bm{a}(t)$ also follows a distribution similar to the one generated by the corresponding PPO.
\end{Rem}

\subsection{Offline training}\label{offlinetraining}
We first simulate an offline environment based on the historical observations on the processes $\{q_n(t)\}$ and $\{g_{n,m,q}(t)\}$. Then, we iteratively generate experiences by interacting with this simulated environment and update the PPO parameters, i.e., $\bm{\theta}_m$ and $\bm{\phi}_m$, based on these experiences.
\subsubsection{Offline environment simulation}\label{offline_env}
We first estimate the distributions of $\{q_n(t)\}$ and $\{g_{n,m,q}(t)\}$ using their historical observations, and then employ these distributions to simulate an offline environment to provide the following functions.
\begin{itemize}
	\item State evolution: We simulate the value of $q_n(t)$ and $g_{n,m,q}(t)$ based on their respective estimated distributions and use them along with the current state $\bm{s}(t)$ to calculate the next state $\bm{s}(t+1)$ and the next agent observations $\{\bm{s}_m(t+1)\}_{m=1}^M$ by using \eqref{eq:qn}, \eqref{eq:cnm}, and \eqref{eq:gnm}.
	\item Reward generation: Reward $r(t)$ is calculated based on \eqref{def:reward}. 
\end{itemize}

\subsubsection{Experience generation phase} During this step, $M$ modified PPOs interact with the offline environment to sequentially generate $N_B$ experiences, where $N_B$ is the size of the experience buffer. Specifically, at the $t$\ts{th} time slot, we send the agent observations $\{\bm{s}_m(t)\}_{m=1}^M$ to $M$ actor networks and derive the action distributions $\{\pi_{m,\bm{\theta}_m}(\cdot|\bm{s}_m(t))\}_{m=1}^M$. We then send these distributions to the DE module to obtain $\bm{a}(t)$ and $\{\pi_{m,\bm{\theta}_m}(a_m(t)|\bm{s}_m(t))\}_{m=1}^M$. Next, we send $\bm{a}(t)$ to the simulated offline environment to obtain reward $r(t)$ and the next set of agent observations $\{\bm{s}_m(t+1)\}_{m=1}^M$. We repeat the above procedures for $N_B$ iterations from the $t$\ts{th} to the $(t+N_B-1)$\ts{th} time slots and pack the related information into $M$ sets of experiences, where the $m$\ts{th} set is denoted as
\begin{align*}
	\{(\bm{s}(t),\bm{s}_m(t),a_m(t),r(t),\pi_{m,\bm{\theta}_m}(a_m(t)|\bm{s}_m(t)))\}_{t}^{t+N_B-1},
\end{align*}
and stored in the experience buffer at PPO$_m$.

\subsubsection{PPO update phase} After generating $N_B$ experiences, we proceed to perform $N_U$ rounds of updates on the parameters of all the $M$ modified PPOs. At the beginning of each update round, we denote the actor and critic networks for PPO$_m$ at that moment as $\pi_{m,\bm{\theta}_m'}$ and $V_{m,\bm{\phi}_m'}$, respectively. Next, we estimate the value functions \cite{ppo}, advantage functions \cite{ppo}, and probability ratios for PPO$_m$ as \cite{liran1}
\begin{align}
    &V_m(t')\!=\!r(t')\!+\!\alpha r(t'\!+\!1)\!+\!\cdots\!+\!\alpha^{N_B-1}r(t\!+\!N_B\!-\!1),\label{eq:ppo1}\\
    &A_m(t')=V_m(t')-V_{m,\bm{\phi}_m'}(\bm{s}(t')),\label{eq:ppo2}\\
    &R_m(t')=\left\{\begin{array}{ll}
    	1&c_m(t')>0\\
    \frac{\pi_{m,\bm{\theta}_m'}(a_m(t')|\bm{s}_m(t'))}{\pi_{m,\boldsymbol{\theta}_m}(a_m(t')|\bm{s}_m(t'))}&c_m(t')=0,
        \end{array}\right.\label{eq:ppo3}
\end{align}
for all $t'\in\{t,t+1,\cdots,t+N_B-1\}$, respectively. Here, the value of probability ratios fluctuates depending on $c_m(t')$, which actually leverages the time-varying constraints in \eqref{con:cmc1} to ensure a faster convergence for PPO$_m$ \cite{liran1}. Then, the surrogate loss for PPO$_m$ is computed as \cite{ppo}
\begin{align}
\begin{split}\label{eq:ppo4}
	L_m=&\sum_{t'=t}^{t+N_B-1}\frac{1}{N_B}\Big(\!-\text{min}\left(R_m(t')A_m(t'),\right.\\
    &\left.\left.\text{clip}(R_m(t'),1-\epsilon,1+\epsilon)A_m(t')\right)\right.\\
    &+c_1A_m(t')^2-c_2H\left(\pi_{m,\bm{\theta}_m'}(\cdot|\bm{s}_m(t'))\right)\!\Big),
\end{split}     
\end{align}
where $\text{clip}(x,a,b)\triangleq\text{min}(\text{max}(x,a),b)$ clamps $x$ into the area $[a,b]$; $H(\pi_{m,\bm{\theta}_m'}(\cdot|\bm{s}_m(t')))$ is the entropy of the distribution $\pi_{m,\bm{\theta}_m'}(\cdot|\bm{s}_m(t'))$; and $\epsilon$, $c_1$, and $c_2$ are some constants. Finally, we use this surrogate loss to update both the actor and critic networks via back propagations \cite{ppo}.

Remarkably, the experience generation phase and PPO update phase are executed alternately during the offline training. The entire offline training algorithm is summarized in Algorithm~\ref{offline}.
\begin{algorithm}[ht]
\caption{Offline training algorithm for multicast scheduling problem {\bf (P1)}}\label{offline}
\begin{algorithmic}[1]
\STATE Initialize the AC networks for $M$ modified PPOs with random parameters $\{\bm{\theta}_m\}_{m=1}^M$ and $\{\bm{\phi}_m\}_{m=1}^M$;
\STATE Define the maximum channel power gain as $L\triangleq\max_{n,m,q,t}g_{n,m,q}(t)$;
\STATE Set the value of the maximum learning episode $L_E$ to a large integer;
\STATE Initialize an experience buffer for each modified PPO;
\STATE Simulate the offline environment according to Section \ref{offline_env};
\STATE \textbf{For} episode$=1,2,\cdots,L_E$ 
\STATE \ \ \ Let $\bm{Q}(1)=0^{N\times M^*}$, $\bm{c}(1)=\bm{0}^{M\times 1}$, $\bm{G}(1)=L^{N\times M}$;
\STATE \ \ \ Derive $\{\bm{s}_m(1)\}_{m=1}^M$ by $\bm{s}_m\!(1)\!=\!(\bm{Q}(1),\!c_m\!(1),\bm{g}_m\!(1))$;
\STATE \ \ \ \textbf{For} $t=1,2,\cdots,N_B$
\STATE \ \ \ \ \ \ Send $\{\!\bm{s}_m(t)\!\}_{m=1}^M$ to $M$ modified PPOs and derive
\STATEx \ \ \ \ \ \ $\{\pi_{m,\bm{\theta}_m}\!(\!\cdot|\bm{s}_m\!(t))\}_{m=1}^M$ based on \eqref{eq:ppo_mod};
\STATE \ \ \ \ \ \ Send $\{\pi_{m,\bm{\theta}_m}(\cdot|\bm{s}_m(t))\}_{m=1}^M$ to DE module and
\STATEx \ \ \ \ \ \ derive $\bm{a}(t)$ and $\{\pi_{m,\bm{\theta}_m}(a_m(t)|\bm{s}_m(t))\}_{m=1}^M$ based
\STATEx \ \ \ \ \ \ on \eqref{eq:de1}, \eqref{eq:de2}, \eqref{eq:de3}, \eqref{anti:con}, and \eqref{eq:de4};
\STATE \ \ \ \ \ \ Send $\bm{a}(t)$ to the simulated offline environment and
\STATEx \ \ \ \ \ \ derive $r(t)$ and $\{\bm{s}_m(t+1)\}_{m=1}^M$ based on \eqref{eq:qn}, \eqref{eq:cnm}, 
\STATEx \ \ \ \ \ \ \eqref{eq:gnm}, and \eqref{def:reward};
\STATE \ \ \ \ \ \ Store $\{(\!\bm{s}(t)\!,\!\bm{s}_m\!(t)\!,\!a_m\!(t)\!,\!r(t)\!,\pi_{m,\bm{\theta}_m}\!(\!a_m\!(t)|\bm{s}_m\!(t))\}\!_{t=1}^{N_B}$
\STATEx \ \ \ \ \ \ in the experience buffer of PPO$_m$;
\STATE \ \ \ \textbf{End for}
\STATE \ \ \ \textbf{For} iteration$=1,2,\cdots,N_U$
\STATE \ \ \ \ \ \ \textbf{For} $m=1,2,\cdots,M$
\STATE \ \ \ \ \ \ \ \ \ Load the stored experiences;
\STATE \ \ \ \ \ \ \ \ \ Calculate the surrogate loss $L_m$ based on \eqref{eq:ppo1},
\STATEx \ \ \ \ \ \ \ \ \ \eqref{eq:ppo2}, \eqref{eq:ppo3}, and \eqref{eq:ppo4};
\STATE \ \ \ \ \ \ \ \ \ Update $\bm{\theta}_m$ and $\bm{\phi}_m$ by backpropagating $L_m$;
\STATE \ \ \ \ \ \ \textbf{End for}
\STATE \ \ \ \textbf{End for}
\STATE \ \ \ Empty all the experience buffers;
\STATE \textbf{End for} 
\end{algorithmic}
\end{algorithm}
\subsection{Online applying}
The online applying algorithm is similar to the offline training one and the main difference between them is that the online one does not store the generated experiences or execute the PPO update in lines 15-21 of Algorithm \ref{offline}. Additionally, the agent observations are derived directly from the real environment, not the simulated one.

\begin{Rem}\label{rem32}
The relationships between conventional DRL {\normalfont\cite{dqn,ddpg}}, the action-embedding method {\normalfont\cite{wolper,yingjun}}, and the proposed distribution-embedding method are summarized as follows.
\begin{itemize}
	\item Conventional DRL algorithms {\normalfont\cite{dqn,ddpg}} directly determine the action value based on the output of their actor networks, while the action-embedding and distribution-embedding methods send the output of the actor network to an additional module to determine the action. By doing so, it is unnecessary to encode the action as a high-dimensional one-hot vector to satisfy the requirement for conventional DRL algorithms, which not only reduces the scale of the NN but also enhances the convergence performance of DRL.
	\item The action-embedding method {\normalfont\cite{wolper,yingjun}} uses an additional module to generate multiple candidate actions based on the suggested action by DRL and then selects the best-evaluated candidate action as the final action. However, this approach faces challenges when dealing with MDPs with large discrete action spaces and time-varying, time-invariant constraints. To elaborate, the method proposed in {\normalfont\cite{wolper}} relies on DDPG training mechanism and generates candidate actions with a poorly interpretive approach. As a result, it exhibits limited convergence performance when applied to MDPs with large discrete action spaces. On the other hand, the method proposed in {\normalfont\cite{yingjun}} achieves promising convergence performance while can only be applied to MDPs with binary-valued high-dimensional action spaces. Moreover, both the methods face challenges when handling MDPs with time-varying or time-invariant constraints. In such cases, their trial-and-error mechanism leads to a high possibility of generating infeasible candidate actions, which hinders the determination of the final action. A possible solution to this issue is to generate a large number of candidate actions, hoping to find feasible ones, while this approach requires an uncertain number of trials and could be problematic for the cases with high-dimensional action spaces. Additionally, if generating one feasible candidate action consumes too many trials, this approach degrades into a purely random policy, rendering the action suggested by the original DRL irrelevant.
    \item The proposed distribution-embedding method can efficiently handle the challenges of time-invariant constraints. It iteratively modifies the action distribution suggested by MADRL and determines action dimension-by-dimension, ensuring that the final action follows an action distribution similar to the one generated by MADRL and satisfies the time-invariant constraints. Finally, by combining the distribution-embedding method with the modified MAPPO, the proposed DE-MAPPO achieves better performance than the existing methods in scenarios with large discrete action space and time-varying, time-invariant constraints.
\end{itemize}
\end{Rem}

\section{Performance Bound}\label{SectionIV}
In this section, we derive an upper bound for the multicast scheduling problem {\bf (P1)}, which is employed as a benchmark to validate the performance of our proposed DE-MAPPO algorithm. To derive the upper bound, we first relax the time-varying constraints \eqref{con:cmc1} and time-invariant constraints \eqref{con:cmc2} of problem {\bf (P1)} to a set of new constraints. Then, we convert the problem under the relaxed constraints to a two-step optimization problem. Finally, we successively solve the optimizations of these two steps, and obtain an upper bound of problem {\bf (P1)}.

\subsubsection{Constraints relaxation} 
We relax constraints \eqref{con:cmc1} and \eqref{con:cmc2} by using the following proposition.
\begin{Prop}\label{prop0}
For any multicast scheduling policy satisfying constraints \eqref{con:cmc1} and \eqref{con:cmc2}, it follows
\begin{align}\label{con:gamma}
	\sum_{n=1}^N\gamma_{n,m}(T)T_{n,m}\leq T,\ m\in\mathcal{M},\ T \in\mathbb{Z}^+,
\end{align}
where $\gamma_{n,m}(T)\in\mathbb{Z}^+$ counts how many times the $n$\ts{th} message has been multicasting over the $m$\ts{th} channel from the $1$\ts{st} to $T$\ts{th} time slots, i.e.,
\begin{align}\label{def:gamma_nm}
\gamma_{n,m}(T)\triangleq\sum_{t=1}^{T}\mathcal{I}_n(a_m(t)),
\end{align}
with $n\in\mathcal{N}$, $m\in\mathcal{M}$, and $T \in\mathbb{Z}^+$.
\end{Prop}
\begin{IEEEproof}[Sketch of proof]
If a multicast scheduling policy satisfies the constraints in \eqref{con:cmc1} and \eqref{con:cmc2}, then all the multicastings of the $n$\ts{th} message over the $m$\ts{th} channel would occupy the $m$\ts{th} channel for $\gamma_{n,m}(T)T_{n,m}$ time slots from the $1$\ts{st} to $T$\ts{th} time slots. As a result, the multicastings of $N$ messages over the $m$\ts{th} channel would occupy the $m$\ts{th} channel for a total of $\sum_{n=1}^N\!\!\gamma_{n,m}(T)T_{n,m}$ time slots from the $1$\ts{st} to $T$\ts{th} time slots, which must not exceed $T$.
\end{IEEEproof}

Proposition \ref{prop0} indicates that constraints \eqref{con:cmc1} and \eqref{con:cmc2} imply constraints \eqref{con:gamma}. Thus, we can simplify problem {\bf (P1)} as
\begin{align*}
    \textbf{(P3)}\ &\max_{\{\bm{a}(t)\}}\ \lim_{T\to\infty}\mathbb{E}_{\{q_n(t)\},\{g_{n,m,q}(t)\}}\left[\frac{1}{T}\sum_{t=1}^{T}r(t)\right]\nonumber\\
&\ \ \text{s.t.}\ \ \ \eqref{eq:qn},\eqref{eq:gnm},\eqref{con:gamma},\nonumber
\end{align*}
where we drop the transitions in \eqref{eq:cnm} since they are no longer needed without constraints \eqref{con:cmc1}.

\subsubsection{Problem conversion}
Now, we convert problem {\bf (P3)} to a two-step optimization problem. In the first step, we fix $\gamma_{n,m}(T)$ by 
\begin{align}\label{con:lastone}
\gamma_{n,m}(T)=\hat{\gamma}_{n,m},
\end{align}
for all $n\in\mathcal{N}$ and $m\in\mathcal{M}$, where $\hat{\gamma}_{n,m}$ is an integer in $\{0,1,\cdots,T\}$. Then, we focus exclusively on optimizing the average latency penalty term of problem {\bf (P3)} while adhering to constraints in \eqref{eq:qn} and \eqref{con:lastone}. To summarize, the first-step optimization for problem {\bf (P3)} is formulated as
\begin{align*}
\textbf{(P4)}\ &\max_{\{\bm{a}(t)\}}-\lim_{T\rightarrow\infty}\mathbb{E}_{\{q_n(t)\}}\left[\frac{1}{T}\sum_{t=1}^{T}\sum_{n=1}^{N}\!\sum_{\tau=1}^{M^*}[\bm{q}_n(t)]_{\tau}p_n(\tau)\right] \nonumber\\
&\ \ \text{s.t.}\ \ \eqref{eq:qn},\eqref{con:lastone}.
\end{align*}
However, solving problem {\bf (P4)} is challenging since it is also a MDP problem with constraints. Therefore, we denote the upper bound of problem {\bf (P4)} as $f(\hat{\boldsymbol{\gamma}})$ with $\left[\hat{\boldsymbol{\gamma}}\right]_{(n,m)}\triangleq \hat{\gamma}_{n,m}$ and use this upper bound to replace the average latency penalty term in problem {\bf (P3)}, which leads to the second-step optimization problem, i.e.,
\begin{align}
\!\!\textbf{(P5)}\!\!&\max_{\{\bm{a}(t)\},\hat{\boldsymbol{\gamma}}}\ -V\!\!\!\lim_{T\to\infty}\!\!\mathbb{E}_{\{\!g_{n,m,q}\!(t)\!\}}\Bigg[\!\frac{1}{T}\!\!\sum_{t=1}^T\!\sum_{n=1}^N\!\sum_{m=1}^{M}\!\!\frac{T_{n,m}Z_{n,m}}{g_{n,m}(t)}\\
&\qquad\qquad \mathcal{I}_n\left(a_m(t)\right)\Bigg]+f(\hat{\boldsymbol{\gamma}})\nonumber\\
&\ \ \ \text{s.t.}\ \ \ \ \eqref{eq:gnm},\nonumber\\
&\ \ \ \ \ \ \ \ \ \ \sum_{n=1}^N\hat{\gamma}_{n,m}T_{n,m}\leq T,\ m\in\{1,2,\cdots,M\},\label{con:p32}
\end{align}
where constraints \eqref{con:p32} are derived from \eqref{con:gamma}. Apparently, the upper bound of problem {\bf (P5)} also serves as an upper bound of problem {\bf (P3)}.

\subsubsection{Upper bound of problem {\bf (P4)}}
To further derive an upper bound of problem {\bf (P4)}, it is observed that the objective function and the constraints of problem {\bf (P4)} can be decomposed into $N$ separate parts, where the $n$\ts{th} part of the objective function is $-\lim_{T\rightarrow\infty}\!\mathbb{E}_{\{q_n(t)\}}\left[\frac{1}{T}\sum_{t=1}^{T}\right.$ $\left.\sum_{\tau=1}^{M^*}[\bm{q}_n(t)]_{\tau}p_n(\tau)\right]$ and the $n$\ts{th} part of the constraint corresponds to the update rule for $\bm{q}_n(t)$. On this basis, we can further decompose problem {\bf (P4)} into $N$ independent sub-problems, where the $n$\ts{th} sub-problem is given as
\begin{align}
\textbf{(P4-1)}\ &\max_{\{a_n(t)\}}\ -\!\!\lim_{T\rightarrow\infty}\!\mathbb{E}_{\{q_n(t)\}}\left[\frac{1}{T}\sum_{t=1}^{T}\sum_{\tau=1}^{M^*}[\bm{q}_n(t)]_{\tau}p_n(\tau)\right]\nonumber\\
&\ \ \text{s.t.}\ \ \ \eqref{eq:qn},\eqref{con:lastone}.\nonumber
\end{align}
Obviously, the sum of the optimal values of $N$ sub-problems {\bf (P4-1)} is the optimal value of problem {\bf (P4)}.

Problem {\bf (P4-1)} can be interpreted as: {\it If the number of the multicastings for the $n$\ts{th} message is fixed as $\sum_{m=1}^M\hat{\gamma}_{n,m}$ within the $1$\ts{st} to $T$\ts{th} time slots, when should we start these multicastings to minimize the average latency penalty?} To answer this question, we let $\pi:\bm{q}_n(t)\rightarrow\{0,1\}$ denote the policy determining whether or not to start a new multicasting at the $t$\ts{th} time slot based on $\bm{q}_n(t)$, and let $t^*$ be the random variable representing the time to start a new multicasting under policy $\pi$ given that the last multicasting occurred at the $0$\ts{th} time slot, i.e., 
$$t^*=\min\{t\in\mathbb{Z}^+|\pi(\bm{q}_n(t))=1,\bm{q}_n(0)=0^{M^*\times 1}\}.$$
Then, problem {\bf (P4-1)} can be reformulated as
\begin{align}
    \textbf{(P4-2)}\ &\max_{\pi:\bm{q}_n(t)\rightarrow\{0,1\}}\ -\mathbb{E}_{\pi,\{q_n(t)\}}\left[\sum_{t=1}^{t^*}\sum_{\tau=1}^{M^*}[\bm{q}_n(t)]_{\tau}p_n(\tau)\right]\nonumber\\
    &\qquad\text{s.t.}\qquad\ \eqref{eq:qn},\bm{q}_n(0)=0^{M^*\times 1},\nonumber\\
    &\qquad\qquad\quad\ \ \mathbb{E}_{\pi,\{q_n(t)\}}t^*=\frac{T}{\sum_{m=1}^M\hat{\gamma}_{n,m}},\label{con:new}
\end{align}
where constraint \eqref{con:new} is derived from \eqref{con:lastone}.

To solve problem {\bf (P4-2)}, we apply the vanilla DQN algorithm \cite{dqn}. Particularly, for the case of $p_n(\tau)\!=\!1$, we have $\sum_{\tau=1}^{M^*}[\bm{q}_n(t)]_{\tau}p_n(\tau)=\sum_{\tau=1}^{M^*}[\bm{q}_n(t)]_{\tau}$, by which we can rewrite problem {\bf (P4-2)} as
\begin{align}
\textbf{(P4-3)}\ &\max_{\pi:\bar{q}_n(t)\rightarrow\{0,1\}}\ -\mathbb{E}_{\pi,\{q_n(t)\}}\left[\sum_{t=1}^{t^*}\bar{q}_n(t)\right]\nonumber\\
&\qquad\text{s.t.}\qquad\ \eqref{con:new},\bar{q}_n(0)=0,\nonumber\\
&\qquad\qquad\quad\ \ \bar{q}_n(t+1)\!=\!\left\{\!\!\begin{array}{ll}
	q_n(t)&b_n(t)=1\\
	\bar{q}_n(t)\!+\!q_n(t)&b_n(t)=0,
\end{array}\right.\!\label{eq:p3_3_2}
\end{align}
where $\bar{q}_n(t)$ is defined as $\bar{q}_n(t)\triangleq\sum_{\tau=1}^{M^*}[\bm{q}_n(t)]_{\tau}$, and \eqref{eq:p3_3_2} is derived from \eqref{eq:qn}. The optimal policy for problem {\bf (P4-3)} can be validated to be of threshold type \cite{huang_stop}, which starts a new multicasting at the $t$\ts{th} time slot as long as $\bar{q}_n(t)$ exceeds certain threshold. Here, the threshold can be calculated based on the optimal stopping rule \cite{huang_stop}.

In summary, by solving problem {\bf (P4-2)} or {\bf (P4-3)}, we derive an upper bound for problem {\bf (P4)}.

\subsubsection{Upper bound of problem {\bf (P5)}} 
Denote the minimum energy consumption to multicast the $n$\ts{th} message over the $m$\ts{th} channel as $e_{n,m}$, i.e.,
\begin{align*}
    e_{n,m}\triangleq\frac{T_{n,m}Z_{n,m}}{\max_{t\in\mathbb{Z}^+}g_{n,m}(t)}.
\end{align*}
Then, based on the definition of $\gamma_{n,m}(T)$ in \eqref{def:gamma_nm}, it follows
\begin{align*}
	&\!-\!V\!\!\!\lim_{T\to\infty}\!\!\mathbb{E}_{\{g_{n,m,q}\!(t)\}}\!\left[\!\frac{1}{T}\!\sum_{t=1}^T\!\sum_{n=1}^N\!\sum_{m=1}^{M}\frac{T_{n,m}Z_{n,m}}{g_{n,m}(t)}\mathcal{I}_n\left(a_m(t)\right)\right]\\
	\leq&\!-\!V\!\!\!\lim_{T\to\infty}\frac{1}{T}\sum_{n=1}^N\sum_{m=1}^Me_{n,m}\gamma_{n,m}(T).
\end{align*}
Therefore, problem {\bf (P5)} can be relaxed as
\begin{align*}
\textbf{(P5-1)}\!\!\!&\max_{\boldsymbol{\gamma}(T)\in\mathbb{N}^{N\times M}}\!-V\!\!\!\lim_{T\to\infty}\frac{1}{T}\sum_{n=1}^N\!\sum_{m=1}^M\!\!e_{n,m}\gamma_{n,m}(T)\!+\!f(\boldsymbol{\gamma}(T))\\
&\ \ \ \ \ \text{s.t.}\ \ \ \ \ \ \eqref{con:gamma},
\end{align*}
where $\boldsymbol{\gamma}(T)$ is defined by $[\boldsymbol{\gamma}(T)]_{(n,m)}\triangleq\gamma_{n,m}(T)$. Apparently, Problem {\bf (P5-1)} can be solved using integer programming \cite{ip_book} or the Lagrange multiplier technique by reforming random variable $\boldsymbol{\gamma}(T)$ as $\bar{\gamma}_{n,m}\triangleq\frac{\gamma_{n,m}(T)}{T}\in[0,1]$, and the derived optimal value of problem {\bf (P5-1)} serves as an upper bounds for problem {\bf (P5)} and {\bf (P3)}.

\begin{figure*}
\centering
\includegraphics[width=6in]{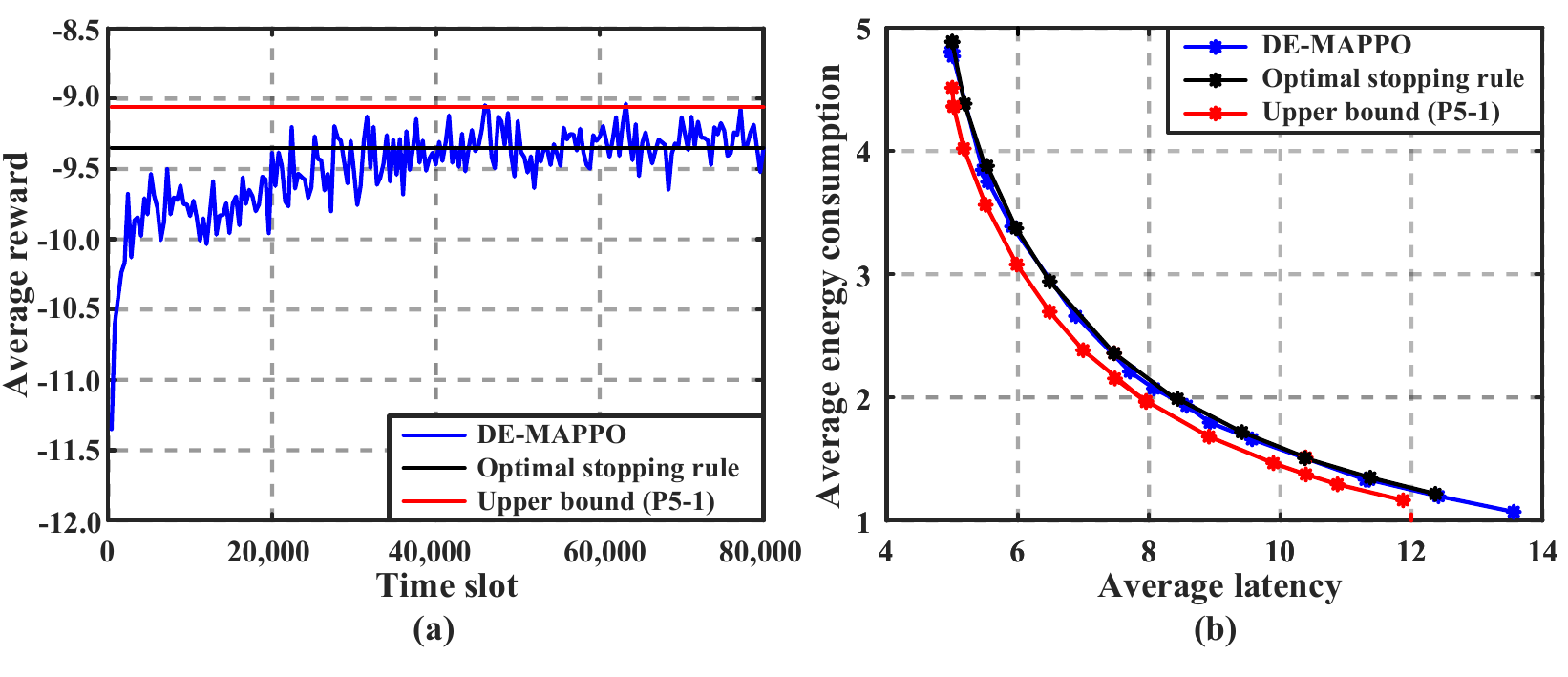}
\caption{Performance comparison among DE-MAPPO, optimal stopping rule, and the upper bound {\bf (P5-1)} for the scenario with $N=1$, $M=1$, $T_{n,m}=1$, and $p_n(\tau)=1$. (a) average reward as a function of time slot in the offline training; (b) average latency vs. average energy consumption tradeoff.} \label{fig1}
\end{figure*}

\section{Simulation Results}\label{SectionV}

This section presents the performance evaluation of the proposed DE-MAPPO algorithm. We employ the following benchmarks:
\begin{itemize}
	\item Optimal stopping rule \cite{huang_stop}: an optimal policy for the case with $N=1$ and $M=1$;
	\item Relative value iteration (RVI) \cite{cui_ite}: an optimal policy solely optimizing the latency penalty for the case with $N=2$ and $M\geq 1$;
	\item MAPPO \cite{liran1}: a state-of-the-art DRL policy applicable for the case with $N\geq 1$ and $M\geq 1$. It can also address the issue of time-varying constraints by utilizing \eqref{eq:ppo_mod} and \eqref{eq:ppo3} in the offline training and online applying phases, while it is not able to handle time-invariant constraints.
	\item Wolpertinger policy (WP) \cite{wolper}: another state-of-the-art DRL policy applicable for the case with $N\geq 1$ and $M\geq 1$. It generates $k$ candidate actions at each time slot and pick the best-evaluated one as the final action;
	\item Performance upper bound derived by solving problem {\bf (P5-1)} (abbreviated as upper bound {\bf (P5-1)});
	\item Round-Robin (RR) \cite{rr}: a typical benchmark policy where $M$ channels take turns to multicast $N$ messages.
\end{itemize}

We set the simulation parameters as follows. The pmf $f_n$ follows Poisson distribution, with its mean value $\lambda_n$ uniformly sampled from the set $\{10,11,\cdots,20\}$. Following a similar setting in \cite{huang_stop}, the downlink channel power gains $g_{n,m,q}(t)$ are uniformly distributed over the range $\{1.00, 1.01, \ldots, 1.10\}$, and the constant $Z_{n,m}$ is set to 5. The size of the request buffer is set as $M^*=4$. In the adopted PPO modules, we set $N_B=1000$, $N_U=10$, $\alpha=0.9$, $\epsilon=0.2$. The learning rate for the actor and critic networks are both set as 0.001.

In Fig. \ref{fig1}, we investigate the performances of DE-MAPPO and the optimal stopping rule, as well as the upper bound {\bf (P5-1)} in the scenario with one message and one channel ($N=1$ and $M=1$). Specifically, we set $T_{1,1}=1$ and $p_1(\tau)=1$ for all $\tau\in\mathbb{Z}^+$, making the optimal stopping rule applicable. In this scenario, DE-MAPPO corresponds to the pure PPO algorithm \cite{ppo}, utilizing two-layer neural networks with 16 nodes for both the actor and critic networks. Fig. \ref{fig1}(a) shows that DE-MAPPO converges after 40,000 time slots and achieves similar average reward compared to the optimal stopping rule method. To evaluate the tradeoff between average latency and average energy consumption achieved by different algorithms, we analyze their performances under various values of $V$ and plot the corresponding tradeoff curves in Fig. \ref{fig1}(b). The results reveal that DE-MAPPO and the optimal stopping rule exhibit coinciding tradeoff curves. Notably, the optimal stopping rule represents an optimal policy, highlighting the promising performance of DE-MAPPO in this scenario.

\begin{figure*}
\centering
\includegraphics[width=6.8in]{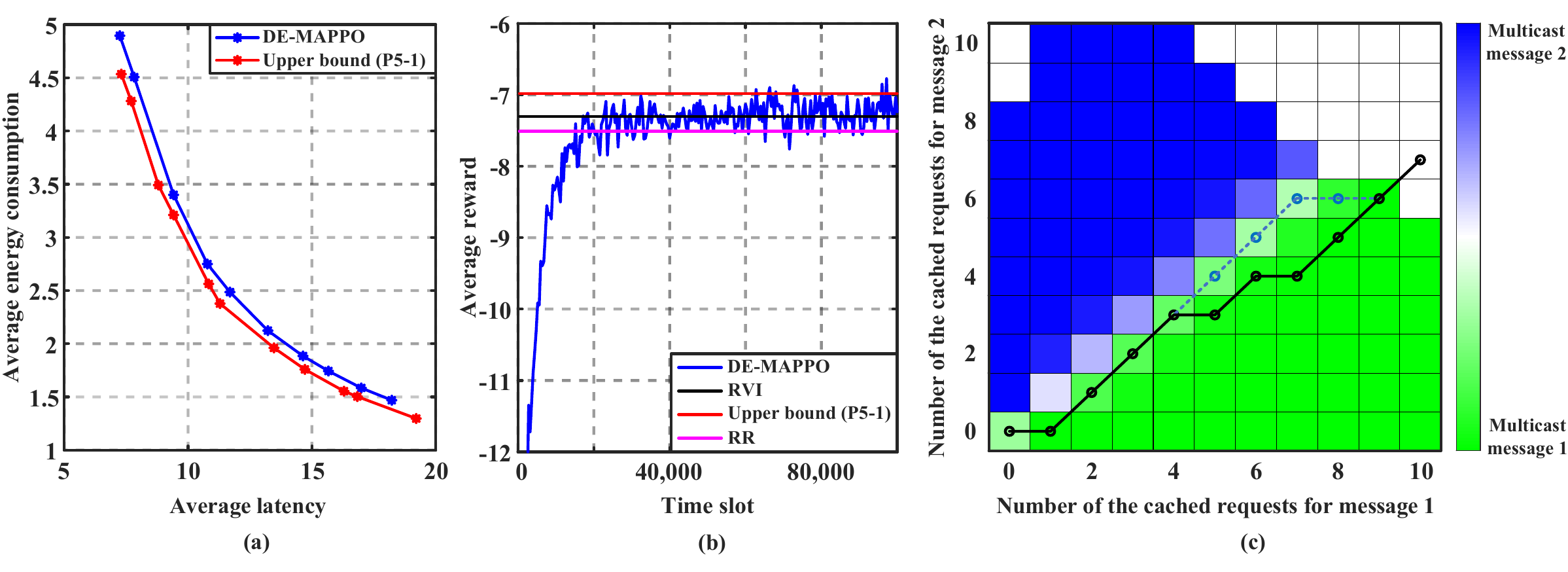}
\caption{Performance comparison among DE-MAPPO, RVI, upper bound {\bf (P5-1)}, and RR for the scenario with $N=2$, $M=1$, $T_{n,m}=1$, $p_n(\tau)=1$, and the size of the request buffers as 10. (a) average latency vs. average energy consumption tradeoff; (b) average reward as a function of time slot in the offline training; (c) scheduling policy structure of DE-MAPPO and RVI.}\label{fig2}
\end{figure*}
Fig. \ref{fig2} investigates the performances of DE-MAPPO, RVI, upper bound {\bf (P5-1)}, and RR in the scenario with two messages and one channel, where the mean request arrival rates for the two messages are set as $\lambda_1=2$ and $\lambda_2=3$, respectively. To make the RVI algorithm applicable, we also limit the size of the request buffer for both messages as 10. The DE-MAPPO adopts two-layer NNs for both the actor and critic networks of the PPO, with each layer containing 32 nodes. Firstly, we evaluate the performance of DE-MAPPO under various values of $V$ in Fig. \ref{fig2}(a), and we observe that DE-MAPPO again achieves a slightly worse performance than upper bound {\bf (P5-1)}. In Fig. \ref{fig2}(b), we observe that DE-MAPPO and RVI achieve similar average reward and outperform RR algorithm. Finally, Fig. \ref{fig2}(c) shows the structure of the scheduling policy for DE-MAPPO and RVI under this scenario. The state of this scenario is a two-dimensional vector containing the numbers of cached requests for the two messages and plotted as squares in Fig. \ref{fig2}(c). At any state, the DE-MAPPO scheduling policy is stochastic and uses a certain Bernoulli distribution to multicast two messages. Specifically, in states colored with deeper blue, DE-MAPPO assigns a higher probability of multicasting message 2, while in states colored with deeper green, it assigns a higher probability of multicasting message 1. In contrast, the optimal policy derived by RVI is deterministic and multicasts message 2 when the state lies above the plotted black curve and message 1 when the state lies on or below the curve. Both DE-MAPPO and RVI showcase a switch-like behavior to determine which message to multicast, and their switch curves exhibit similarities. Note that RVI is an optimal policy under the investigated scenario. The observed similarities in performance between DE-MAPPO and RVI provide strong evidence for the effectiveness of DE-MAPPO.

\begin{figure*}
\centering
\includegraphics[width=6.8in]{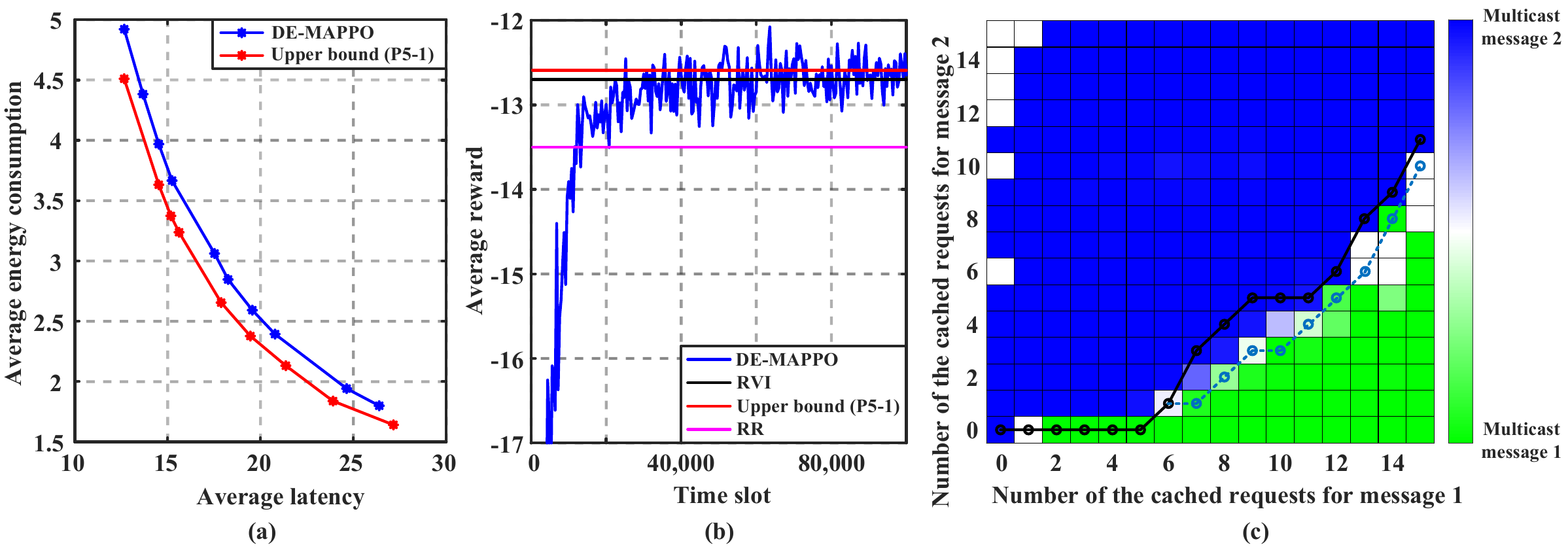}
\caption{Performance comparison among DE-MAPPO, RVI, upper bound {\bf (P5-1)}, and RR for the scenario with $N=2$, $M=1$, $T_{n,m}=1$, $p_n(\tau)=1$, and the size of the request buffers as 15. (a) average latency vs. average energy consumption tradeoff; (b) average reward as a function of time slot in the offline training; (c) scheduling policy structure of DE-MAPPO and RVI.}\label{fig3}
\end{figure*}
Fig. \ref{fig3} investigates the performances of the algorithms in a more complex scenario, where the mean request arrival rates for the two messages are $\lambda_1=2$ and $\lambda_2=7$, and the request buffer size for both messages is 15. The results obtained by DE-MAPPO are similar to those in Fig. \ref{fig2}, demonstrating its ability to achieve high rewards and a switch-like scheduling policy similar to RVI. However, it should be noted that obtaining the optimal policy using RVI requires evaluating the relative value over 256 states, which is a computationally intensive process. Moreover, the computational requirements would exponentially increase when extending RVI to the scenario with larger request buffer sizes or message numbers. In contrast, as shown in Fig. \ref{fig2}(b) and Fig. \ref{fig3}(b), the proposed DE-MAPPO can achieve comparable rewards to RVI with reasonable offline training times, which involve around 250 and 300 times of PPO parameter update, respectively. This highlights the advantages of DE-MAPPO over RVI in terms of computational efficiency.

\begin{figure*}
\centering
\includegraphics[width=6in]{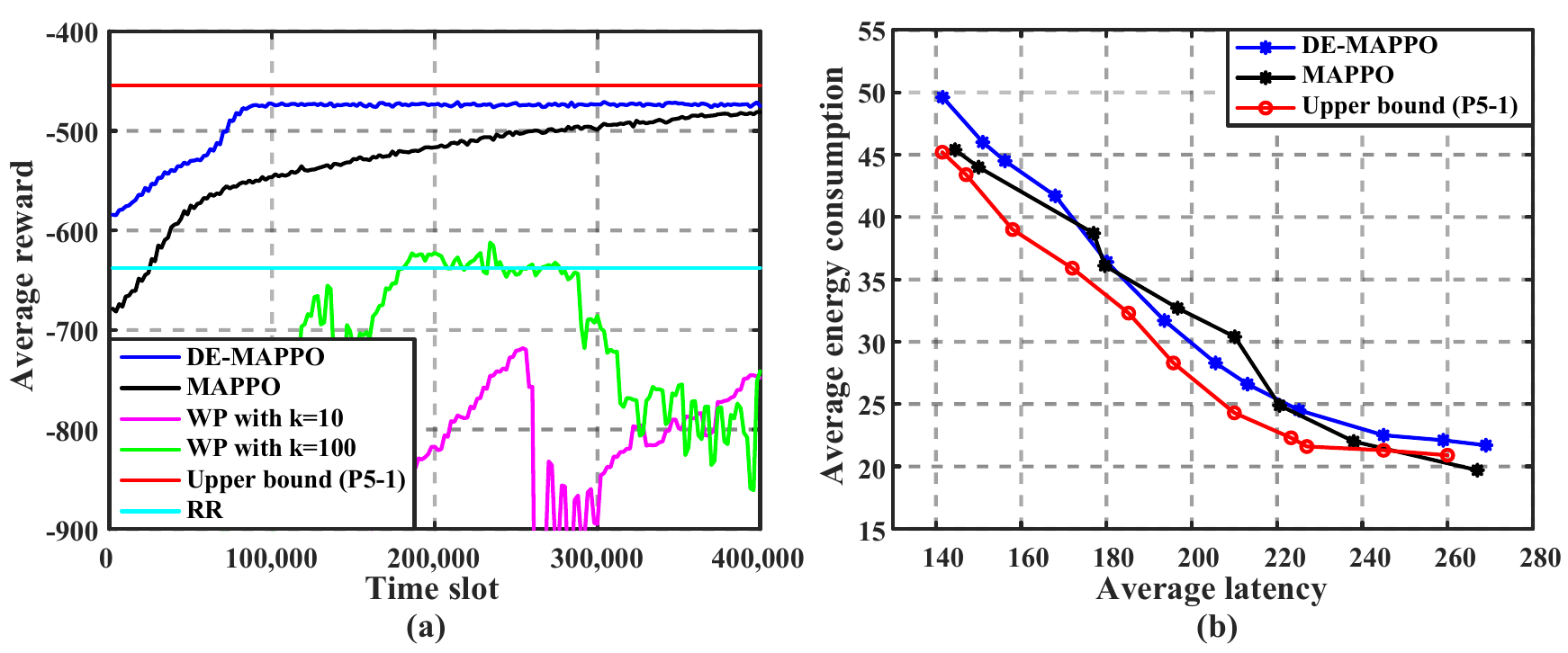}
\caption{Performance comparison among DE-MAPPO, MAPPO, WPs, upper bound {\bf (P5-1)}, and RR for the scenario with $N=10$, $M=10$, $T_{n,m}=1$, and $p_n(\tau)=1$. (a) average reward as a function of time slot in the offline training; (b) average latency vs. average energy consumption tradeoff.} \label{fig4}
\end{figure*}
Fig. \ref{fig4} examines the performances of the algorithms in the scenario with 10 messages and 10 channels ($N=10$ and $M=10$), where the action space has a cardinality of $11^{10}$ and is too large for optimal stopping rule and RVI. In this setting, we compare DE-MAPPO with MAPPO, WP with $k=10$ and $k=100$, upper bound {\bf (P5-1)}, and RR algorithm. All DRL algorithms use three-layer NNs for both the actor and critic networks with 128 nodes per layer. As shown in Fig. \ref{fig4}(a), DE-MAPPO converges faster than MAPPO and achieves a similar average reward. The average reward of DE-MAPPO is also close to that of upper bound {\bf (P5-1)} and much higher than that of RR algorithm. WP algorithms do not converge well in this scenario because they cannot effectively address MDPs with large action spaces, as highlighted in Remark \ref{rem32}. Moreover, MAPPO is not able to restrict itself to selecting only feasible actions that satisfy the time-invariant constraints in \eqref{con:cmc2}. Consequently,  it optimizes the average reward within a larger action space than the original MDP and thus may achieve a higher average reward than the optimal policy of the original MDP or even upper bound {\bf (P5-1)}. However, due to the same reason, MAPPO is actually not applicable in practice. Fig. \ref{fig4}(b) shows that DE-MAPPO has a tradeoff curve close to MAPPO and upper bound {\bf (P5-1)}, indicating its effectiveness in balancing the average latency and average energy consumption. These results demonstrate that DE-MAPPO is capable of efficiently solving MDPs with large discrete action spaces and time-invariant constraints.

\begin{figure*}
\centering
\includegraphics[width=6in]{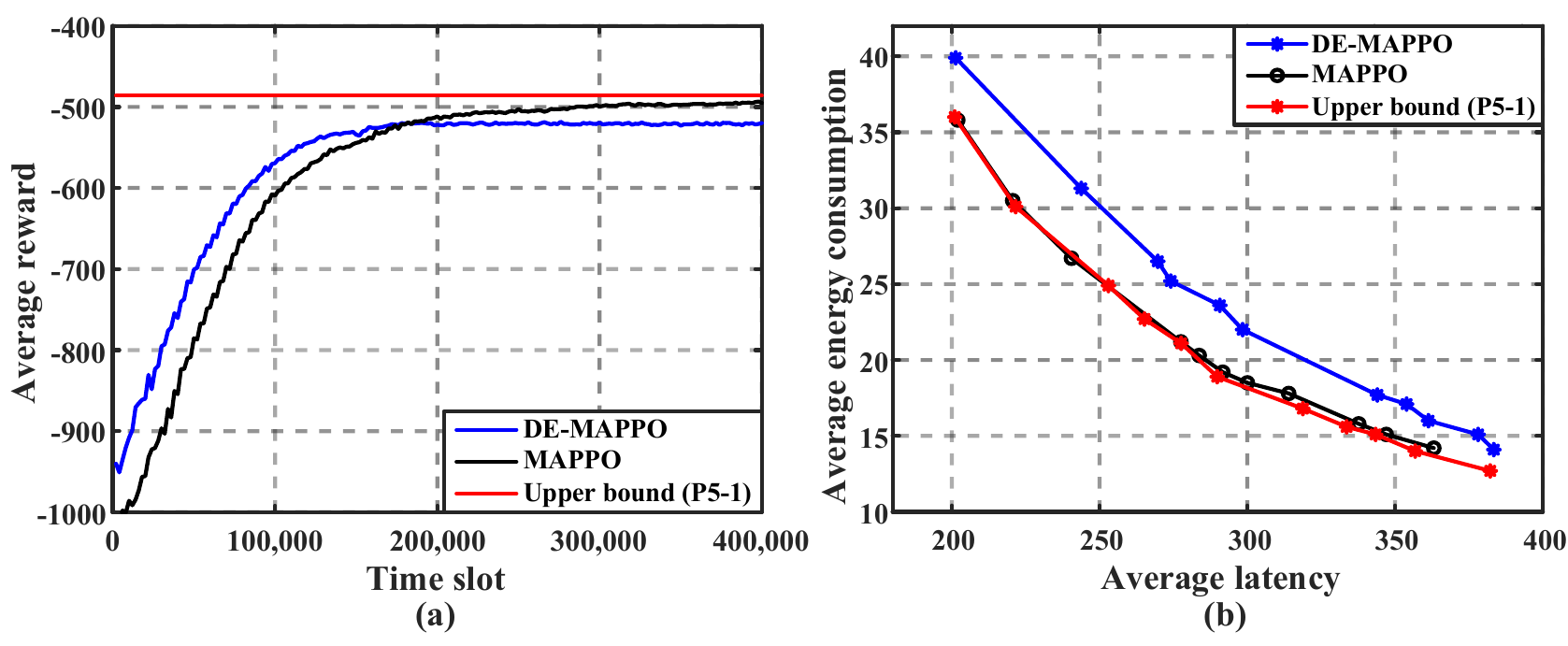}
\caption{Performance comparison among DE-MAPPO, MAPPO, and upper bound {\bf (P5-1)} for the scenario with $N=10$, $M=10$, $T_{n,m}\in\{1,2,\cdots,5\}$, and $p_n(\tau)=1$. (a) average reward as a function of time slot in the offline training; (b) average latency vs. average energy consumption tradeoff.}\label{fig5}
\end{figure*}
\begin{figure*}
\centering
\includegraphics[width=6in]{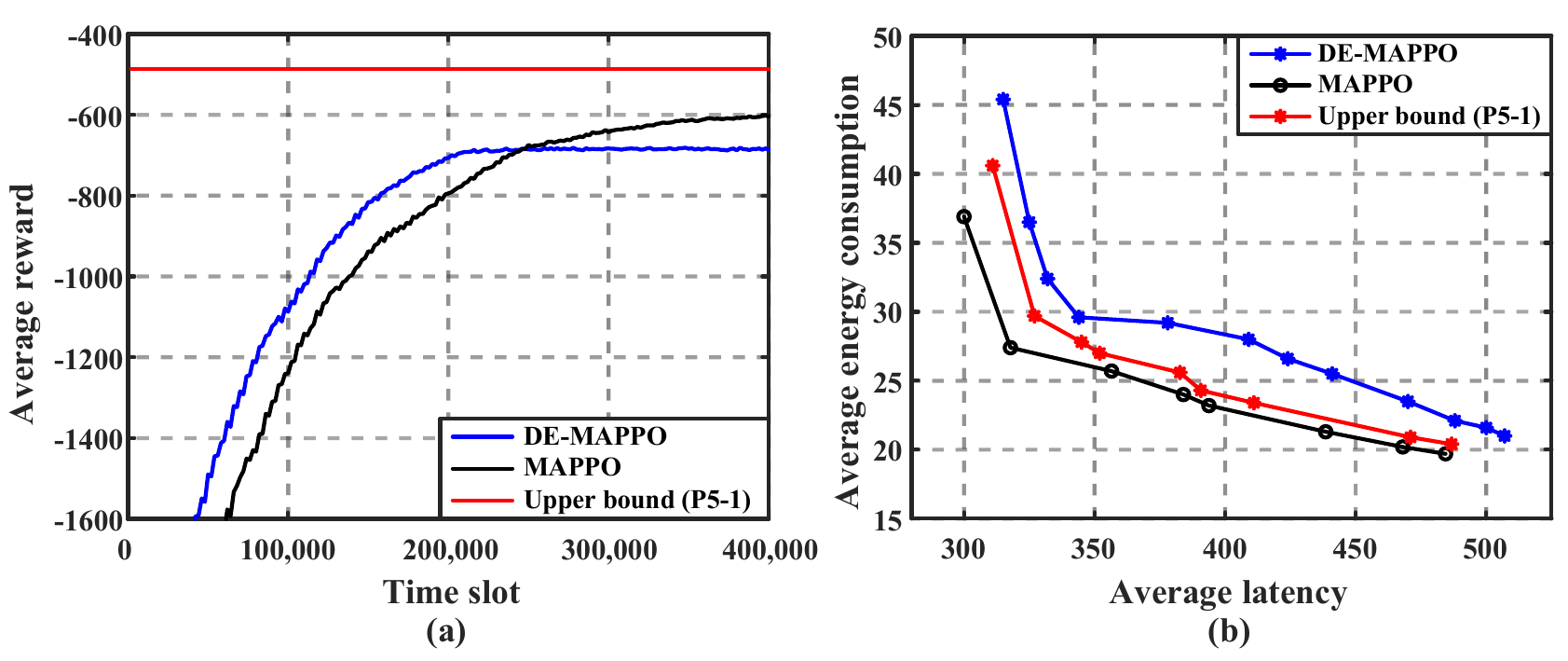}
\caption{Performance comparison among DE-MAPPO, MAPPO, and upper bound {\bf (P5-1)} for the scenario with $N=10$, $M=10$, $T_{n,m}\in\{1,2,\cdots,5\}$, and $p_n(\tau)=\tau$. (a) average reward as a function of time slot in the offline training; (b) average latency vs. average energy consumption tradeoff.}\label{fig6}
\end{figure*}
In Fig. \ref{fig5}, we investigate the performances of the algorithms in a more complex scenario, where the multicasting of each message over each channel may consume up to 5 time slots, i.e., $T_{n,m}\in\{1,2,\cdots,5\}$. In this scenario, the scheduling decision at each time slot must also satisfy the time-varying constraints in \eqref{con:cmc1}. We evaluate various algorithms in Fig. \ref{fig5}(a), where WP and RR algorithms are not plotted since their performances are too bad. DE-MAPPO again converges faster than MAPPO and achieves comparable average reward to MAPPO and upper bound {\bf (P5-1)}. In Fig. \ref{fig5}(b), the tradeoff curve of DE-MAPPO is again close to those of MAPPO and upper bound {\bf (P5-1)}. These results demonstrate the superiority and effectiveness of DE-MAPPO when dealing with MDPs characterized by large discrete action spaces and both time-varying and time-invariant constraints.

In Fig. \ref{fig6}, we consider the case of 10 messages and 10 channels with a linear latency penalty function $p_n(\tau)=\tau$. The results are similar to those in Fig. \ref{fig5}, which demonstrates the effectiveness of DE-MAPPO in scenarios with a general latency penalty function.

\section{Conclusions}\label{SectionVI}
We consider the multicast scheduling problem for multiple messages over multiple channels, which jointly optimizes the average energy consumption and the average latency penalty. This problem is formulated as an infinite-horizon MDP, which is challenging due to the large discrete action space, multiple time-varying constraints, and multiple time-invariant constraints. To address the first two issues, a modified MAPPO utilizing adapted actor networks and enhanced training kernel is adopted. To address the time-invariant issue, a novel distribution-embedding approach is proposed, which iteratively modifies the output distribution of MAPPO and outperforms the existing action-embedding approaches. Finally, a performance upper bound is derived by solving a two-step optimization problem. Numerical results demonstrate that our proposed algorithm outperforms the existing algorithms and is comparable to the derived benchmark.

\bibliographystyle{IEEEtran}%
\bibliography{ref.bib}

\begin{IEEEbiography}[{\includegraphics[width=1in,clip,keepaspectratio]{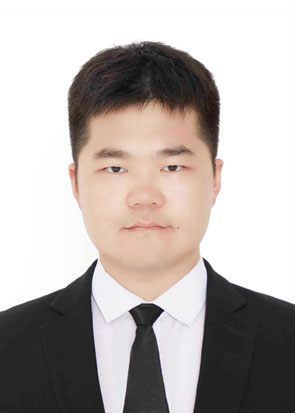}}]{Ran Li} (Graduate Student Member, IEEE) received the B.E. degree in communication engineering from University of Electronic Science and Technology of China (UESTC), Chengdu, China, in 2017. He is currently pursuing the Ph.D. degree with the School of Science and Engineering (SSE), The Chinese University of Hong Kong, Shenzhen, China. His current research interests include reinforcement learning and resource allocation in wireless networks.
\end{IEEEbiography}

\begin{IEEEbiography}[{\includegraphics[width=1in,clip,keepaspectratio]{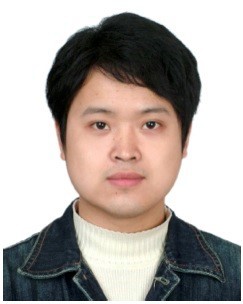}}]{Chuan Huang} (Member, IEEE) received the Ph.D. degree in electrical engineering from Texas A\&M University, College Station, USA, in 2012. From August 2012 to July 2014, he was a Research Associate and then a Research Assistant Professor with Princeton University and Arizona State University, Tempe, respectively. He is currently an Associate Professor with The Chinese University of Hong Kong, Shenzhen. His current research interests include wireless communications and signal processing. He served as a Symposium Chair for IEEE GLOBECOM 2019 and IEEE ICCC 2019 and 2020. He has been serving as an Editor for IEEE TRANSACTIONS ON WIRELESS COMMUNICATIONS, IEEE ACCESS, Journal of Communications and Information Networks, and IEEE WIRELESS COMMUNICATIONS LETTERS.
\end{IEEEbiography}

\begin{IEEEbiography}[{\includegraphics[width=1in,clip,keepaspectratio]{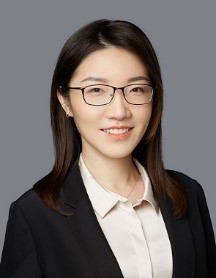}}]{Xiaoqi Qin} (Member, IEEE) received her B.S., M.S., and Ph.D. degrees from Electrical and Computer Engineering with Virginia Tech. She is currently an Associate Professor of School of Information and Communication Engineering with Beijing University of Posts and Telecommunication(BUPT). Her research focuses on exploring performance limits of next-generation wireless networks, and developing innovative solutions for intelligent and efficient machine-type communications.
\end{IEEEbiography}

\begin{IEEEbiography}[{\includegraphics[width=1in,clip,keepaspectratio]{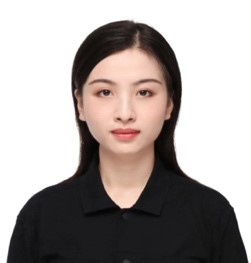}}]{Xinyao Nie} received her B.S. degree from School of Computer with Central China Normal University (CCNU) and M.S. degree from Computer Science and Technology with Fudan University (FDU). She is currently a computer vision engineer in SF Technology. Her research focuses on digital twin and computer vision in logistics.
\end{IEEEbiography}
\vspace{-18cm}
\begin{IEEEbiography}[{\includegraphics[width=1in,clip,keepaspectratio]{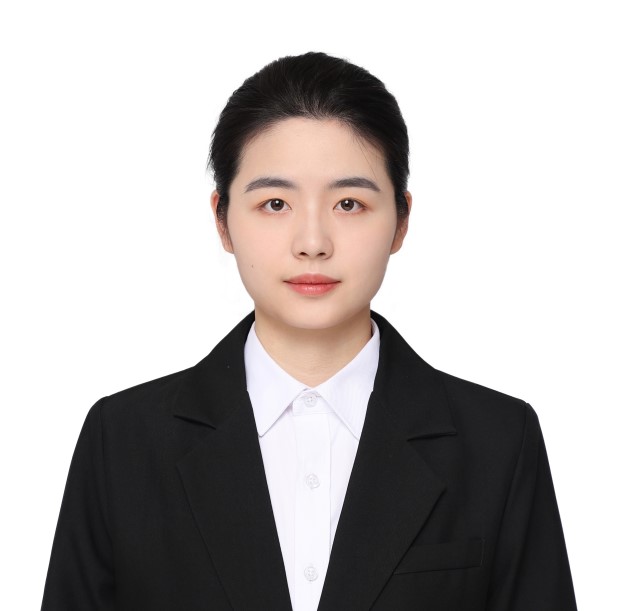}}]{Zhanhong Fu} received her B.S. degree from Computer Engineering with Universit\'{e} de Lorraine and M.S. degree from Intelligent Systems Engineering with Universit\'{e} Paris VI. She is currently a computer vision engineer in SF Technology. Her research focuses on digital twin and multimodal large language model.
\end{IEEEbiography}

\end{document}